\documentclass[12pt]{article}

\usepackage{scicite}
\usepackage[font=small,labelfont=bf]{caption}

\usepackage{times}

\usepackage{amsmath}
\usepackage{amssymb}
\usepackage{graphicx}
\usepackage{dcolumn}
\usepackage{bm}
\usepackage[rightcaption]{sidecap}
\usepackage{hyperref}
\usepackage{xcolor}
\usepackage{upgreek}
\usepackage{lscape}

\newenvironment{sciabstract}{
\begin{quote} \bf}
{\end{quote}}

\title{\Large{\textbf{Multi-scale dynamics of\\colloidal deposition and erosion in porous media }}}

\author{\normalsize{Navid Bizmark,$^{1,2}$ Joanna Schneider,$^{2}$ Rodney D. Priestley,$^{1,2}$ Sujit S. Datta$^{2\ast}$}\\

\normalsize{%
 $^{1}$ Princeton Institute for the Science and Technology of Materials,}\\
 \normalsize{%
 Princeton University, 	
Princeton, NJ 08544
}\\

\normalsize{%
 $^{1}$ Department of Chemical and Biological Engineering,}\\
 \normalsize{%
 Princeton University, 	
Princeton, NJ 08544
}\\

\normalsize{$^\ast$To whom correspondence should be addressed; E-mail:  ssdatta@princeton.edu.}\\
}

\date{}

\begin{document}

\baselineskip24pt

\maketitle 

\begin{sciabstract}\small{Diverse processes---\textit{e.g.}, environmental pollution, groundwater remediation, oil recovery, filtration, and drug delivery---involve the transport of colloidal particles in porous media. Using confocal microscopy, we directly visualize this process \textit{in situ} and thereby identify the fundamental mechanisms by which particles are distributed throughout a medium. At high injection pressures, hydrodynamic stresses cause particles to be continually deposited on and eroded from the solid matrix---strikingly, forcing them to be distributed throughout the entire medium. By contrast, at low injection pressures, the relative influence of erosion is suppressed, causing particles to localize near the inlet of the medium. Unexpectedly, these macroscopic distribution behaviors depend on imposed pressure in similar ways for particles of different charges, even though the pore-scale distribution of deposition is sensitive to particle charge. These results reveal how the multi-scale interactions between fluid, particles, and the solid matrix control how colloids are distributed in a porous medium.}\\

\small{Teaser: Microscopy reveals that the interplay of deposition and erosion controls the distribution of colloids throughout a porous medium.}

\end{sciabstract}

\newpage\section*{Introduction}

Colloidal transport underlies a wide array of processes that impact our everyday lives. It can be beneficial, helping to degrade contaminants in groundwater aquifers \cite{Phenrat2009,Zhao2016,Kanel2006}, mobilize oil from underground reservoirs \cite{Zhang2014}, and carry therapeutics through gels and tissues in the body \cite{Tang2013,Vlashi2013}. It can also be harmful, enabling the distribution of microplastics, contaminants, and pathogens throughout soils, sediments, and groundwater aquifers \cite{Schijven2003,Zhong2017,Harvey1991}. Being able to predict and control colloidal transport is therefore critically important. However, while transport is well-studied in bulk liquids, all of these examples involve the transport of colloidal particles in a disordered, three-dimensional (3D) porous medium. In this case, not only do confinement and tortuosity imposed by the medium alter particle transport, but the particles in turn can alter the medium by depositing onto its solid matrix as they are transported \cite{Datta1998a,Datta1998b,Sahimi1991,Zeman2017,Linkhorst2016}, yielding coupled dynamics that pose a challenge to current understanding. 

Indeed, even basic studies of particle transport and deposition often yield conflicting results. For example, in many cases, colloids have been found to enhance the transport of chemicals through a porous medium \cite{Weber2009,Kersting1999,Ryan1996}, while other studies report the opposite, claiming that colloids suppress transport by reducing the permeability of the medium \cite{Tiraferri2008,Civan2010,Wiesner1996,Liu1995}. This discrepancy is thought to be rooted in the different flow conditions and colloidal chemistries explored, which can strongly influence the interactions between flowing fluid, particles, and the medium. However, systematic investigation of these interactions remains challenging; typical 3D media are opaque, precluding direct characterization of flow and particle transport within the pore space. Thus, experiments often investigate flow around isolated obstacles \cite{Kusaka2010,Charvet2017} or through two-dimensional (2D) arrays of pores \cite{Lin2016,Auset2006,Wyss2006,RobertdeSaintVincent2016,Lin2017}. While these studies provide tremendous insight into particle transport and deposition, they do not fully capture the connectivity and complexity of a 3D pore space. Recent work has extended these investigations to transparent 3D media, but only investigated single particle behavior \cite{Schwartz2019}, did not investigate different colloidal chemistries \cite{Mays2011,Mays2015}, or only focused on particle deposition near the inlet of the medium \cite{Gerber2019}. Further, magnetic resonance imaging and X-ray microtomography yield additional insights into the spatial distribution of particles in the pore space---for example, indicating that the distribution of deposited particles is sensitive to the particle charge \cite{Li2006a,Li2006b}, size \cite{Gerber2018}, and imposed flow conditions \cite{Li2005}. However, such approaches typically do not capture pore-scale dynamics of colloidal transport and deposition due to limitations in spatial or temporal resolution.

In lieu of direct microscopic observations, studies often use a combination of continuum modeling and filtration theory \cite{Molnar2015} to describe particle transport and deposition. In this hybrid approach, a single parameter---the collector efficiency---quantifies all particle interactions with the medium (\textit{e.g.}, due to electrostatics). While in many cases this approach provides a useful way to model colloidal transport and deposition, it often does not reliably predict experimentally observed deposition profiles without the use of additional fitting parameters, reflecting the combined influence of many different factors on particle transport and deposition \cite{Molnar2015,Messina2016,Boccardo2018}. To shed further light on this problem, various computational schemes are being developed, generating intriguing predictions of particle deposition and erosion \cite{Hilpert2018,Yang2017,Jager2017,Pham2017,Sanaei2018,Ding2015,Miele2019}. However, in the absence of experimental studies connecting the dynamics of colloidal deposition and erosion at the pore scale to flow and transport at the scale of the overall porous medium, accurate prediction or control of particle transport and deposition remains elusive. 

Here, we address this gap in knowledge by directly visualizing the dynamics of colloidal particle transport in transparent, 3D porous media. Our experiments probe length scales ranging from single pores to thousands of pores and time scales ranging from the injection duration of one pore volume to thousands of pore volumes, enabling us to connect pore-scale dynamics to macroscopic transport. As particles are transported through a pore space, they deposit on the surrounding solid matrix of the medium. At high injection pressures, hydrodynamic stresses also continually erode the particles from the solid matrix, causing deposited particles to be distributed through the entire porous medium. Conversely, at low injection pressures, hydrodynamic stresses are weaker and the relative influence of erosion is suppressed, causing deposited particles to be localized only near the inlet of the medium. Unexpectedly, these macroscopic distribution behaviors are tuned by the elevated imposed pressure in similar ways for particles of different charges, even though the pore-scale distribution of deposited particles is sensitive to particle charge. Thus, our results reveal the rich pore-scale dynamics of colloidal transport in a porous medium, deepening understanding of how the multi-scale interactions between flowing fluid, particles, and the solid matrix impact colloidal transport and deposition.\\

\section*{Results} 
\subsection*{Experimental platform for visualization of colloidal dynamics} 
\begin{figure}
    \centering
    \includegraphics[width=0.9\textwidth]{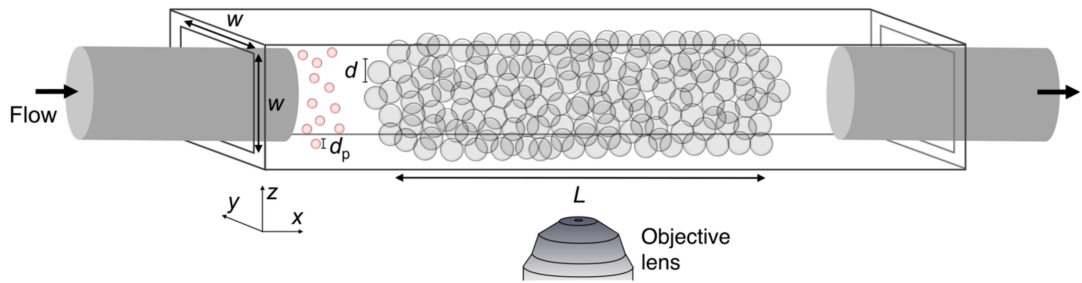}    
    \caption{\textbf{Schematic of experimental setup.} Our experiments are performed using 3D porous media composed of glass beads of diameter $d$ between $38$ and $45~\upmu$m, densely packed within thin-walled square quartz capillaries with cross-sectional length $w = 1$ mm. A dilute colloidal suspension containing particles of diameter $d_{\text{p}} = 1~\upmu$m is injected through the pore space at a fixed imposed pressure drop. The fluorescent particles are visualized at scales ranging from that of individual pores to the overall porous medium using confocal microscopy.}
    \label{fig1}
\end{figure}

We prepare rigid 3D porous media by lightly sintering dense, disordered packings of hydrophilic glass beads, with diameters $d$ between $38$ and $45~\upmu$m, in thin-walled square quartz capillaries of cross-sectional area $A$ = $1$ mm$^2$. The experimental geometry is schematized in Figure \ref{fig1}. The packings have lengths $L$ ranging from $5$ mm to $2$ cm and porosity $\phi_{0}\approx0.41$ as measured previously using confocal microscopy \cite{Datta2013}. Scattering of light from the surfaces of the beads typically precludes direct observation of flow and transport within the medium. Following our previous work \cite{Datta2013,Johnson2013}, we overcome this limitation by formulating a fluid mixture with a refractive index matching that of the glass beads—enabling full characterization of pore space structure and subsequent visualization of colloidal deposition using confocal microscopy (\textit{Materials and Methods}). 

Prior to each experiment, we map the pore space by saturating the medium with the fluorescently dyed fluid and acquiring a cross-sectional image of the full pore space at a fixed depth in the medium. We identify the glass beads by their contrast with the dyed fluid, with cross-sections shown by the black circles in Figure \ref{fig2}, and the fluid-saturated pore space by the bright region between the beads. Thus, this protocol enables us to characterize the pore space structure at sub-pore resolution prior to colloidal injection.  

We then inject the undyed fluid, laden with fluorescent colloidal particles of diameter $d_{\text{p}} = 1 ~\upmu$m at a concentration of $\sim10^9$ particles/mL = $0.05$ vol\%, at a fixed pressure drop $\Delta P$ across the overall medium. The magnitude of the characteristic interstitial flow velocity ranges from $\sim0.5$ to $5$ cm/min, or $\sim7$ to $70$ m/day, comparable to that of forced-gradient groundwater flow in a sand aquifer. We denote the time $t$ at which particles begin to enter the medium as $t = 0$, and represent subsequent times by the total number of suspension pore volumes (PVs) injected, $\int_{0}^{t}Q(t)dt/\left[AL\phi(t)\right]$, where $Q(t)$ and $\phi(t)$ are the time-dependent volumetric flow rate and porosity, respectively. To characterize particle deposition in the pore space, we continually acquire successive images spanning the entire cross-section of the medium; in parallel, we measure the effluent mass, providing a direct measure of the flow rate. The Reynolds number characterizing our experiments is $\sim10^{-3}$ to $10^{-4}$, indicating that the flow is laminar. The particle P\'eclet number Pe, quantifying the importance of advection relative to diffusion in determining the particle motion, is $>10^4$; hence, particle transport is primarily due to fluid advection. Because our experiments explore the injection of up to thousands of pore volumes of particle suspension, they range from ``clean bed" conditions characterized by minimal prior deposition to nearly clogged conditions. Further, our experiments test two different colloidal particle chemistries characterized by ``favorable" attractive electrostatic interactions or ``unfavorable" repulsive interactions with the solid matrix of the medium. Thus, our work complements previous characterization of these distinct conditions and modes of interaction using transport measurements, magnetic resonance imaging, X-ray microtomography, optical imaging, and static light scattering \cite{Zeman2017,Linkhorst2016,Li2005,Gerber2019,Li2006a,Li2006b,Mays2011,Mays2015,Gerber2018,Schwartz2019}. 

\subsection*{Dynamics of positively-charged colloidal particles}
\begin{figure}
    \centering
    \includegraphics[width=\textwidth]{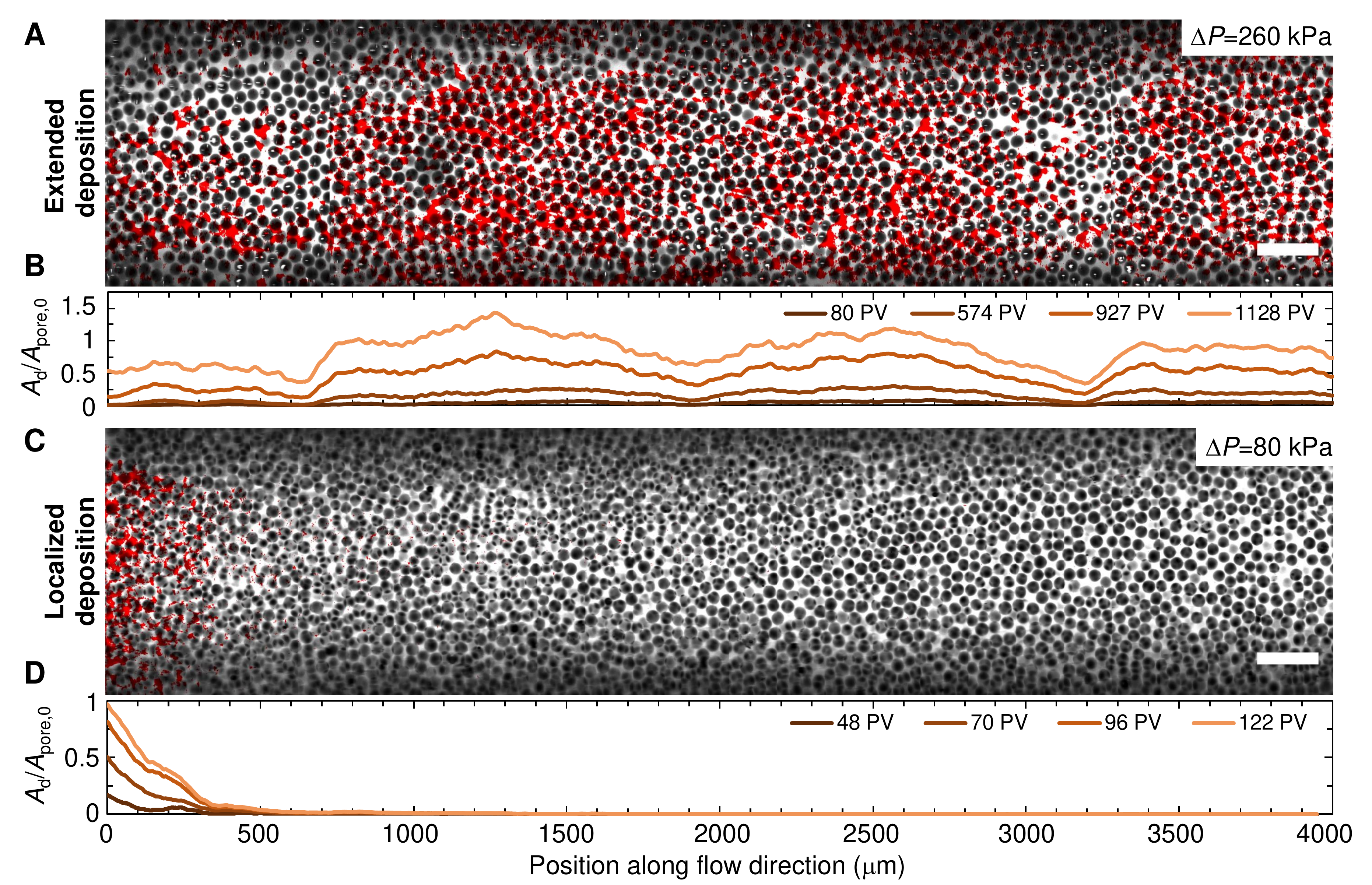}    
    \caption{\textbf{Macroscopic deposition profile of positively-charged colloidal particles is tuned by injection pressure.} Confocal micrographs show extended (A) and localized (C) deposition profiles after $t = 3.1$ h and $1.9$ h for amine-functionalized polystyrene particles injected at $260$ or $80$ kPa, respectively. Black circles show cross-sections through the beads making up the porous media, white space shows pore space, and red shows deposited colloidal particles. Corresponding line traces in (B) and (D) show the amount of deposition at each position along the flow direction, quantified by $A_{\text{d}}/A_{\text{pore},0}$ where $A_{\text{d}}$ is the laterally-averaged area occupied by deposited particles and $A_{\text{pore},0}$ is the pore space area before colloidal injection. Traces show $A_{\text{d}}$ with a running average over every $100~\upmu$m along the length of the medium to minimize noise; different traces are shown for images obtained every $2$ min in (B) and $3$ min in (D), which we represent by the total number of suspension pore volumes (PVs) injected. The traces exhibit some spatial fluctuations that likely reflect the influence of packing heterogeneities in the disordered media. In some cases, $A_{\text{d}}/A_{\text{pore},0}$ slightly exceeds one due to the non-zero thickness of the optical slice. For clarity, we only show the deposition profile for the first $4000~\upmu$m of the media, shorter than their full lengths. Flow direction is from left to right and the scale bars represent $200~\upmu$m.}
    \label{fig2}
\end{figure}

We first investigate the injection of a dilute suspension of positively-charged, amine-functionalized polystyrene particles into the porous medium at a large imposed pressure drop, $\Delta P = 260$ kPa. In this case, we anticipate particle deposition to be localized near the inlet of the medium due to strong electrostatic attraction between the positively-charged particles and negatively-charged beads \cite{Ryan1996,Johnson2013}, as suggested by the DLVO calculations detailed in the \textit{Supplementary Materials} and summarized in Figure S1. Surprisingly, we instead observe the formation of an extended deposition profile of particles that spans the entire length of the porous medium, as shown in Figure \ref{fig2}A and Movie S1. This profile persists over the course of injection: the amount of deposition at each position along the flow direction---quantified by the fraction of the initial pore space area $A_{\text{pore},0}$ occupied by deposited particles having total area $A_{\text{d}}$---steadily grows in time, as shown in Figures \ref{fig2}B and S2. 

Inspection of particle deposition at the pore scale provides a clue to the mechanism underlying extended deposition. Because Pe $\gg1$ and particle-bead interactions are attractive, we expect fluid flow to advect particles towards the upstream faces of the beads, forcing them to initially attach. Consistent with this expectation, some particles initially deposit on the upstream surfaces of the pristine beads, as shown in the top panel of Figure \ref{fig3}A. As particle injection continues, these deposits continue to grow, as shown in Movie S2---suggesting that the hydrodynamic stresses due to fluid flow are sufficient to overcome any possible electrostatic repulsion between like-charged particles, detailed further in the \textit{Supplementary Materials}, although nanoscale heterogeneity in the colloidal interactions may also play a role \cite{VanNess2019}. This observation is consistent with previous findings in 2D media \cite{Lin2016,Auset2006,Wyss2006,RobertdeSaintVincent2016,Lin2017,Gerber2019}. The deposits do not all grow monotonically, however. In many cases, single- and multi-particle deposits are eroded—--abruptly removed by fluid flow, as predicted by recent simulations \cite{Jager2017}; an example is shown in the bottom panel of Figure \ref{fig3}A and in Movie S3. The eroded particles are then re-deposited in other pores downstream. This process of deposition, erosion, and re-deposition progresses continuously over the course of injection, enabling particles to be deposited throughout the extent of the medium. Moreover, while significant, erosion does not completely balance deposition: the instantaneous rate of pore-scale deposition is slightly larger than the rate of erosion, as shown by the solid points in Figure \ref{fig3}B. This difference results in net deposition throughout the entire medium that progressively increases over time, as quantified in the top panel of Figure \ref{fig3}C. Therefore, we hypothesize that pore-scale erosion generates the extended macroscopic deposition profile shown in Figure \ref{fig2}A.

To quantify this hypothesis, we analyze the stresses on colloidal particles at the pore scale (\textit{Materials and Methods}). Because the imposed pressure drop is so large, we expect that hydrodynamic stresses are large enough to drive erosion. As a first step toward quantifying this expectation, we use Darcy’s law to estimate the characteristic viscous pressure drop across an isolated deposit of diameter $D$, $\Delta P_{\text{D}}=\mu QD/\left(Ak\right)$, at $t=0$; here, $\mu$ is the fluid dynamic shear viscosity and $k$ is the permeability of the porous medium, measured previously for a pristine medium \cite{Krummel2013}. For deposits of size $D$ ranging from one to four particle diameters, as seen in Movie S3, $\Delta P_{\text{D}}$ ranges from $\sim10$ to $60$ Pa. We conjecture that this characteristic pressure drop is much larger than the yield stress required to fluidize dense colloidal aggregates with slight interparticle attraction, $\sigma_{y}$, which we estimate to be between $\sim1$ and $10$ Pa based on previous shear rheology measurements \cite{Pham2008}. Hence, we expect that fluid flow drives erosion, as observed in the experiments. This balance of pressures neglects spatiotemporal variations in flow as well as the full details of colloid-colloid and colloid-bead interactions; it therefore provides an order-of-magnitude estimate of the onset of erosion. To more rigorously test our expectation, we perform mechanistic particle trajectory simulations explicitly incorporating the particle forces and torques that arise from the interplay between hydrodynamics and colloidal interactions \cite{VanNess2019}. The simulations confirm our expectation: they reveal that multi-particle clusters are both deposited and eroded from the surfaces of the beads under these conditions, as shown in Figure S3 and detailed in the \textit{Supplementary Materials}. As particles continue to deposit, the permeability of the medium and the volumetric flow rate decrease, as shown in the middle panel of Figure \ref{fig3}C. The interstitial flow speed $v(t)\equiv Q(t)/\left[\phi(t)A\right]$, calculated by directly estimating the time-dependent porosity $\phi(t)\equiv \phi_{0}\left[1–A_{\text{d}}(t)/A_{\text{pore},0}\right]$ from our micrographs, concomitantly decreases as shown in the bottom panel of Figure \ref{fig3}C; this quantity provides a measure of the magnitude of the characteristic interstitial flow velocity. However, even at these lower flow rates and flow speeds, $\Delta P_{\text{D}}\gtrsim\sigma_{y}$, enabling erosion to continue to progress. Thus, particles continue to deposit, erode, and re-deposit in the entire porous medium over the course of the experiment, driving extended deposition throughout.

Our analysis also suggests that the relative contribution of erosion can be tuned by the imposed pressure drop. To test this prediction, we repeat our experiments, but at a lower $\Delta P = 80$ kPa. Under these conditions, we expect hydrodynamic stresses and hence the relative magnitude of erosion to be less dominant, likely resulting in different deposition behavior. Consistent with this expectation, we indeed observe starkly different deposition behavior: instead of particles depositing throughout the entire porous medium, as in the high pressure case, particles only deposit locally near the inlet, as shown in Figure \ref{fig2}C and Movie S4. We quantify this observation in Figure \ref{fig2}D, which shows that the fraction of the pore space that is occupied by particles no longer spans the entire length of the medium, but instead appears to decay exponentially, as shown in Figure S2, approaching zero at a distance $L_{\text{d}}\sim7\%$ of its length. The deposition profiles at different times collapse when rescaled by the number of injected particles, as shown in Figure S2, indicating that the deposition process is consistent over time \cite{Gerber2018,Johnson2018,Johnson2020}.

As expected, these differences in macroscopic deposition reflect strong differences in pore-scale deposition and erosion, in addition to previously-documented differences in particle transport at different pressures \cite{Li2005}. For $\Delta P = 80$ kPa, the instantaneous rates of particle deposition and erosion---as well as the difference between them---are one order of magnitude larger than at $\Delta P = 260$ kPa, as shown by the open points in Figure \ref{fig3}B. This difference results in a larger fraction of particles depositing in the medium over a shorter amount of time, shown in the top panel of Figure \ref{fig3}C. As a result, the volumetric flow rate decreases, indicating that localized deposition at the inlet ``chokes off" the flow, as shown in the middle panel of Figure \ref{fig3}C. The corresponding interstitial flow speed is nearly an order of magnitude smaller than that of the high pressure case, as shown in the bottom panel of Figure \ref{fig3}C. Consequently, the characteristic viscous pressure drop across a deposit, $\Delta P_{\text{D}}\sim 1$ to $6$ Pa, is insufficient to strongly overcome the deposit yield stress, estimated as $\sigma_{y}\sim1$ to $10$ Pa, and erosion is suppressed. Mechanistic particle trajectory simulations of particle dynamics confirm this expectation, indicating that erosion is suppressed under these conditions, as shown in Figure S4. Therefore, particle deposition is localized to near the inlet of the medium. \\

\subsection*{Dynamics of negatively-charged colloidal particles} 
Our experiments thus far explored the case of positively-charged particles. How does deposition change when the particle charge is altered? To answer this question, we next investigate the injection of a dilute suspension of negatively-charged carboxyl-functionalized polystyrene particles. Intriguingly, at a large $\Delta P = 170$ kPa, we again observe the formation of an extended deposition profile spanning the length of the porous medium, as shown in Figures \ref{fig4}A and S2, and in Movie S5. However, microscopy reveals that pore-scale deposition is dramatically altered by the change in particle charge. In this case, because the particles have the same charge as the beads composing the medium, the electrostatic interactions between particles and beads are repulsive. Hence, carboxyl-functionalized particles do not deposit on the upstream surfaces of the pristine beads; instead, they are strained in the tight pore throats between beads, where they continue to grow and form loose deposits, exemplified by Figure \ref{fig5}A. This observation corroborates similar findings in experiments using larger particles \cite{Li2006a,Li2006b}. 

Despite this difference in where particles deposit, we observe similar deposition dynamics to the amine-functionalized case: some deposits grow monotonically, while in many other cases, single- and multi-particle deposits are eroded and re-deposited downstream, as shown in the bottom panel of Figure \ref{fig5}A and in Movies S6 and S7. Erosion does not completely balance deposition: the instantaneous rate of pore-scale deposition is slightly larger than the rate of erosion, indicated by the solid points in Figure \ref{fig5}B. This difference again results in net deposition in the entire medium that progressively increases over time, as shown in the top panel of Figure \ref{fig5}C. The volumetric flow rate and interstitial flow speed, shown in the middle and bottom panels of Figure \ref{fig5}C, respectively, decrease concomitantly. However, hydrodynamic stresses are still sufficient to strongly drive erosion: the characteristic viscous pressure drop across a deposit, $\Delta P_{\text{D}}\sim10$ to $60$ Pa, is larger than the deposit yield stress, again estimated as $\sigma_{y}\sim1$ to $10$ Pa. Particle trajectory simulations of particle dynamics again confirm this expectation, revealing that clusters of particles are both deposited and eroded from the surfaces of the beads under these conditions, as shown in Figure S5. Therefore, a similar process of continual flow-driven deposition, erosion, and re-deposition again results in extended deposition of particles throughout the medium. 

Consistent with this picture, at a smaller $\Delta P=80$ kPa, we again find that deposition is localized---in this case, to the first $\sim20$\% of the length of the medium, as shown in Figure \ref{fig4}C and Movie S8. The deposition profiles at different times also collapse when rescaled by the number of injected particles, as shown in Figure S2, indicating that the deposition process is consistent over time \cite{Gerber2018,Johnson2018,Johnson2020}. At this lower pressure, the instantaneous rates of particle deposition and erosion---as well as the difference between them---are over one order of magnitude larger than at $\Delta P=170$ kPa, shown by the open points in Figure \ref{fig5}B. This difference again results in a larger fraction of particles depositing in the medium over a shorter amount of time for the lower pressure drop condition, shown in the top panel of Figure \ref{fig5}C. The volumetric flow rate decreases concurrently, resulting in slow interstitial flow that cannot drive erosion as strongly, as shown by the middle and bottom panels of Figure \ref{fig5}C, respectively: the characteristic viscous pressure drop $\Delta P_{\text{D}}\sim1$ to $6$ Pa is much smaller than in the $170$ kPa case, and is insufficient to overcome the deposit yield stress, estimated as $\sigma_{y}\sim1$ to $10$ Pa. Mechanistic particle trajectory simulations of particle dynamics additionally confirm this expectation, indicating that erosion is suppressed under these conditions, as shown in Figure S6. Together, our experiments reveal that macroscopic deposition of colloidal particles is tuned by imposed pressure in similar ways, independent of particle charge.\\

\subsection*{Changes in the overall permeability of the medium} 
Having established how hydrodynamics impact macroscopic deposition of particles, we now ask how deposition in turn impacts fluid flow. Our observations of localized deposition---primarily in a region spanning a length $L_{\text{d}} < L$ along the porous medium---suggest that the overall permeability of the medium can be calculated by considering flow in the deposited and pristine, particle-free regions separately. In particular, for a given time $t > 0$ after the initiation of particle injection, we use Darcy’s law to describe the pressure drop across the overall medium as 
\begin{equation}
    \begin{aligned}
    \Delta P={
  \underbrace{\frac{\mu Q(t)L_{\text{d}}(t)}{A k_{\text{d}}(t)}}%
    _{\text{Deposited region}}
 }\,+ {
  \underbrace{\frac{\mu Q(t)\left[L-L_{\text{d}}(t)\right]}{A k_{0}}}%
    _{\text{Pristine region}}
 }\,
    \label{eq1}
    \end{aligned}
\end{equation}
\noindent where $k_{\text{d}}$ and $k_0$ are the permeabilities of the particle-deposited and pristine particle-free regions, respectively; in the case of extended deposition, $L_{\text{d}} = L$ in this equation. This pressure drop can also be related to the initial volumetric flow rate $Q_0$ before the initiation of particle injection: $\Delta P = \mu Q_{0}L/\left(Ak_{0}\right)$. Because $\Delta P$ is fixed in our experiments, combining this equation with Eq. \ref{eq1} yields a prediction for the overall permeability of the medium, $k$: 
\begin{equation}
    \begin{aligned}
    \tilde{k}(t)\equiv \frac{k(t)}{k_{0}}=\tilde{Q}(t)\equiv \frac{Q(t)}{Q_{0}}=\frac{1}{1+\tilde{L}_{\text{d}}(t)\left[1/\tilde{k}_{\text{d}}(t)-1\right]},
    \label{eq2}
    \end{aligned}
\end{equation}
\noindent where $\tilde{L}_{\text{d}}(t)\equiv L_{\text{d}}(t)/L$ and $\tilde{k}_{\text{d}}(t)\equiv k_{\text{d}}(t)/k_{0}$ are the normalized deposition length and permeability of the deposit-filled region, respectively. As shown in Figure \ref{fig6}A, this prediction quantifies the intuition that the overall permeability of a medium with minimal colloidal deposition ($\tilde{L}_{\text{d}}\ll1$, $\tilde{k}_{\text{d}}\approx1$) is only minimally altered, while a medium with substantial deposition ($\tilde{L}_{\text{d}}\approx1$, $\tilde{k}_{\text{d}}\ll1$) becomes clogged and impermeable to further flow. Thus, processes that seek to control the permeability of the medium during colloidal injection should focus on controlling the two parameters $\tilde{L}_{\text{d}}$ and $\tilde{k}_{\text{d}}$.

Our experiments enable a direct test of this prediction: our transport measurements shown in Figures \ref{fig3}C and \ref{fig5}C directly yield $\tilde{k}(t)=\tilde{Q}(t)$, while our macroscopic and pore-scale visualization yield $\tilde{L}_{\text{d}}(t)$ and $\tilde{k}_{\text{d}}(t)$, respectively, enabling us to independently compute $\tilde{k}(t)$ \textit{via} Eq. \ref{eq2}. To determine $\tilde{L}_{\text{d}}(t)$, we use the micrographs shown in Figures \ref{fig2} and \ref{fig4} to determine the position along the medium at which $A_{\text{d}}/A_{\text{pore},0}$ falls below a threshold value, as detailed in the \textit{Supplementary Materials}. To determine $\tilde{k}_{\text{d}}(t)$, we use the confocal micrographs to determine the pore space area $A_{\text{pore}}(t)$, the pore space perimeter $P_{\text{pore}}(t)$, and the porosity $\phi(t)$, all of which are time-dependent. We then estimate $\tilde{k}_{d}(t)$ by modeling the pore space as a parallel bundle of capillary tubes (\textit{Materials and Methods}): $\tilde{k}_{\text{d}}(t)=\left[\phi(t)/\phi_{0}\right]\left[d_{\text{h}}(t)/d_{\text{h,0}}\right]^{2}$, where the hydraulic diameter $d_{\text{h}}(t)\sim A_{\text{pore}}(t)/P_{\text{pore}}(t)$. Finally, we compute $\tilde{k}(t)$ \textit{via} Eq. \ref{eq2}. Remarkably, despite the simplifying assumptions made, this model yields reasonable agreement---within a factor of $\sim2$---between the measured and computed $\tilde{k}$ for all of the previously described experiments, as well as additional experiments performed at intermediate pressure drops and higher inlet particle concentrations, as shown in Figure \ref{fig6}B. By connecting colloidal deposition at the pore scale to changes in macroscopic transport, this model helps to confirm the consistency of our data.

\section*{Discussion}
Our work directly connects the dynamic processes of colloidal deposition and erosion at the pore scale, deposition profiles at the macroscopic scale, and bulk fluid transport. Importantly, we find that particles can deposit throughout the entire medium at large pressures \textit{via} continual erosive bursts. Erosion was previously theorized to occur using pore-scale simulations \cite{Jager2017}, and indirect signatures of erosion have been detected using bulk transport measurements \cite{Bianchi2018}; our results provide a connection between these pore-scale events and macroscopic deposition behavior. Moreover, while the pore-scale characteristics of deposition depend on the interactions between particles and the solid matrix, we find that the macroscopic characteristics of deposition are tuned by imposed pressure in unexpectedly similar ways for particles with different surface properties---highlighting the importance of hydrodynamic interactions in determining colloidal transport and deposition. Specifically, our results suggest that erosion plays a dominant role in distributing particles throughout the pore space when the viscous pressure drop $\Delta P_{\text{D}}$ across deposited particles exceeds a threshold value, which we conjecture is given by the deposit yield stress $\sigma_{y}$. The transition between localized and extended deposition is therefore likely to be abrupt, tuned by the imposed pressure drop, unlike the flow rate-controlled case that has been found to yield more gradual deposition behavior \cite{Gerber2019}. Elucidating this transition will be a valuable direction for future work. Further, it will be interesting to explore whether collective clogging effects \cite{Gerber2018} manifest for both positively- and negatively-charged particles, given the similarity of our results for both cases.

To facilitate visualization using confocal microscopy, our experiments employ an aqueous refractive index-matched mixture as the fluid phase. However, though this fluid mixture has different hydrodynamic, dielectric, and refractive properties than pure water---resulting in differences in colloidal interactions---we expect that our results are also applicable to pure water-based colloids. This expectation is supported by the DLVO calculations shown in Figure S1, which indicate that despite the quantitative differences in colloidal interactions between the two different fluid phases, the nature of colloid-colloid and colloid-bead interactions are qualitatively similar. This expectation is further confirmed by mechanistic particle trajectory simulations explicitly incorporating the different particle forces and torques in a pure water-based system; we find similar behaviors as those obtained using the refractive index-matched fluid mixture, but at different values of the imposed fluid pressure, as detailed in the \textit{Supplementary Materials} and summarized in Figures S3 to S6. Hence, our results are likely generalizable to a broader range of aqueous-based colloids.

Because diverse applications require control over colloidal deposition in porous media, we anticipate our findings will be broadly relevant. In some cases, localized deposition---which we find arises at low injection pressures---is essential. One important example is the deposition of iron nanoparticles within groundwater aquifers for \textit{in situ} immobilization of heavy metal contaminants \cite{Bianco2017}; in this case, nanoparticle deposition is required, but only at specific locations of the medium. Another example is filtration or containment of pathogens and waste materials, which requires effective capture near the inlet of the medium \cite{Schijven2003}. By shedding light on the conditions under which the relative influence of erosion is promoted or suppressed, our work may assist in the determination of optimal injection pressures and medium lengths to promote localized deposition in these cases. In other cases, particles must be able to traverse long distances as they are advected through the pore space. Key examples are enhanced oil recovery \cite{Zhang2014}, in which injected particles must be able to reach residual oil in a reservoir; nanoparticle-assisted groundwater remediation \cite{Phenrat2009,Zhao2016,Kanel2006}, in which injected particles must be able to reach trapped contaminants in an aquifer; and drug delivery through tissues in the body \cite{Tang2013,Vlashi2013}. Our results suggest that, even if particles deposit in the pore space, erosion at sufficiently large injection pressures can be leveraged to distribute the particles through the entire medium. Thus, our work may provide guidelines for more effective colloidal transport in environmental, energy, and biomedical settings.\\

\section*{Materials and Methods}
\noindent\textbf{Experimental setup.} We prepare rigid 3D porous media by lightly sintering dense, disordered packings of hydrophilic glass beads, with diameters $d$ between $38$ and $45~\upmu$m, in thin-walled square quartz capillaries of cross-sectional area $A = 1$ mm$^2$ for under a minute at $900^{\circ}$C. The packings have lengths $L$ ranging from $5$ mm to $2$ cm and porosity $\phi_{0}\approx0.41$ as measured previously using confocal microscopy \cite{Datta2013}. Prior to each experiment, the pore space is saturated with the particle-free fluid before the homogenized colloidal suspension is injected into the medium. We impose a constant pressure drop across the medium using a \textit{Teledyne} ISCO LC-5000 syringe pump, and use an \textit{Omega} differential pressure sensor to independently verify the pressure drop across the medium. We continually image the pore space, as detailed below; in parallel, we measure the volumetric flow rate by measuring the effluent mass over time using a \textit{Mettler Toledo} balance with an accuracy of 1 mg and converting to a volume using a density $\rho\approx1.23$ g/cm$^3$, calculated as a weighted average of the densities of the different fluid components. The experiments last until a filter cake of particles begins to form at the inlet of the medium. \\

\noindent\textbf{Fluid and colloid properties.} We formulate an index-matched aqueous fluid composed of $82$ wt\% glycerol (\textit{Sigma-Aldrich}), $12$ wt\% dimethyl sulfoxide (\textit{Sigma-Aldrich}), and $6$ wt\% ultrapure water. This fluid has a density $\rho\approx1.23$ g/cm$^3$, calculated as a weighted average of the densities of the different fluid components, and a dynamic shear viscosity $\mu\approx60$ mPa-s, as previously determined using a shear rheometer \cite{Datta2013}. The colloids used are fluorescent amine- (\textit{Sigma-Aldrich}) and carboxyl-functionalized polystyrene particles (FluoSpheres, \textit{Thermo Fisher Scientific}) with a mean diameter of $d_{\text{p}} = 1~\upmu$m; DLVO analysis of the colloidal interactions are given in the \textit{Supplementary Materials}. Stock suspensions are sonicated to uniformly disperse the particles, diluted to $0.05$ vol\% (for the experiments described in Figures \ref{fig2} to \ref{fig5}) or 0.1 vol\% (for additional experiments shown in Figure \ref{fig6}) in the fluid mixture, and further homogenized by vortexing and additional sonication. The zeta potentials of the amine- and carboxyl-functionalized particles are $+25\pm8$ and $-34\pm5$ mV, respectively, as measured in the fluid mixture itself using a \textit{Malvern} Zetasizer Nano-ZS. We verify that the magnitudes of these zeta potentials are sufficiently large to maintain particle stability in the suspension over the entire course of injection, as detailed in the \textit{Supplementary Materials}. We also do not observe any noticeable deformation or swelling of the polystyrene particles in the presence of glycerol and DMSO at the ratios used in our tests; this observation is consistent with previous reports indicating that polystyrene does not swell or soften when exposed to these solvents \cite{Iseda2011,Lumma2000,ThermoFisher}. The Reynolds number characterizing our experiments is $\rho v d_{\text{b}}/\mu \leq 3\times10^{-4}$ where $d_{\text{b}}$ is the diameter of a pore body, indicating that the flow is laminar. The particle P\'eclet number Pe $= 6\pi\mu v(d_{\text{p}}/2)^{2}/\left(k_{\text{B}}T\right) \sim 10^{4}$ to $10^5$, where $k_{\text{B}}$ is Boltzmann's constant and $T$ is temperature, and the ratio of viscous forces to gravitational forces on the particles is given by $9\mu v/\left[2\Delta \rho g(d_{\text{p}}/2)^{2}\right] \sim 10^5$ to $10^6$, where $\Delta\rho\approx0.17$ g/cm$^3$ is the density difference between particles and fluid and $g$ is gravitational acceleration---indicating that particle transport is primarily due to advection by the fluid.\\

\noindent\textbf{Confocal microscopy.} Prior to each experiment, we saturate the pore space with the particle-free fluid, dyed with Rhodamine 6G (\textit{Sigma-Aldrich}). We use a laser scanning Nikon A1R+ confocal microscope to acquire high-resolution optical slices at a fixed depth within the pore space; these span the entire width and length of the medium and thus provide a full cross-sectional image. The acquisition time for the entire cross-section is $\sim2$ min. We identify the glass beads by their contrast with the dyed fluid, with cross-sections shown by the black circles in Figure \ref{fig2}, and the fluid-saturated pore space by the bright region between the beads. Hence, this protocol enables us to characterize the pore space structure prior to colloidal injection.  We then inject the colloidal suspension, composed of fluorescent particles dispersed in the same but undyed fluid, and acquire successive cross-sectional images of particles in the pore space. We adjust laser power and gain amplitude to avoid pixel saturation for each experiment. The time required to raster across the width of each optical slice is longer than the time required for individual particles to be advected across; as a result, the images primarily reflect deposited particles, not particles dispersed in the suspension. The focal depth in each experiment is $7~\upmu$m except for tests with the carboxyl-functionalized polystyrene particles at $80$ kPa, where focal depth is $38~\upmu$m due to the use of a lower-magnification objective lens to scan the longer porous medium; the pore space area $A_{\text{pore}}$ and area of deposited particles $A_{\text{d}}$ determined from the images thus correspond to a two-dimensional projection of a volume spanning $\sim1$ pore body to $\sim1$ bead in depth. For all experiments, we restrict our analysis of the images to a distance $\sim60~\upmu$m away from the transverse boundaries of the capillary to minimize edge effects.  \\

\noindent\textbf{Simulations of particle deposition and erosion.} To explore the influence of hydrodynamic and colloidal interactions on particle deposition and erosion, we use the Parti-Suite software package \cite{PartiSuite} to examine the trajectories of single particles or clusters of particles as they flow through the pore space. Specifically, we simulate Lagrangian trajectories that explicitly incorporate the forces and torques on particles \cite{VanNess2019} in a Happel sphere-in-cell model of the pore space surrounding an individual bead \cite{Molnar2015}. Previous work has established the ability of this framework to quantitatively model particle deposition and erosion from surfaces under the influence of imposed flow \cite{VanNess2019}. Specifically, the particles are initialized at random points at the entrance of the pore upstream of the bead. We choose the flow velocity to match that of the experiments. Prior to particle-bead contact, the particle velocity is computed from the action of hydrodynamic forces exerted by the imposed flow, computed using the parameter values given in Table S2. Upon contact with the bead surface, particle movement is dictated by the balance between hydrodynamic and attachment torques \cite{VanNess2019}, computed using the parameter values given in Tables S1 and S2, respectively. Multi-particle clusters are treated as larger $4~\upmu$m-diameter particles and are simulated in the same way, but using physico-chemical parameters that describe the relevant colloid-colloid interactions. We use these simulations to assess deposition and erosion of amine- and carboxyl-functionalized particles and particle clusters, at high and low imposed pressures, and in both the refractive index-matched fluid mixture and in pure water. These results are detailed in the \textit{Supplementary Materials}.\\

\noindent\textbf{Calculation of permeability of deposit-filled region.} To determine $\tilde{k}_{\text{d}}(t)\equiv k_{\text{d}}/k_{0}$, we use the images in a square region near the inlet of the medium approximately $400~\upmu$m $\times~400~\upmu$m across to determine the time-dependent pore space area $A_{\text{pore}}(t)$, the deposited particle cross-sectional area $A_{\text{d}}(t)$, the pore space perimeter $P_{\text{pore}}(t)$, and the porosity $\phi(t)$ in the particle deposit-filled region. We then calculate $\tilde{k}_{\text{d}}(t)$ by modeling the pore space as a parallel bundle of cylindrical capillary tubes, each of which has a hydraulic diameter $d_{\text{h}}$ and length $l$. The pressure drop across each tube is then given by the Hagen-Poiseuille equation as $32\mu lv/d_{\text{h}}^{2}$, where $v=Q/\left(\phi A\right)$ is the interstitial flow speed in the medium. This pressure drop is also given by Darcy's law as $\mu v \phi l/k_{\text{d}}$. Equating these two relations for pressure drop yields $k_{\text{d}}=\phi d_{\text{h}}^{2}/32$. Hence, $\tilde{k}_{\text{d}}(t)=\left[\phi(t)/\phi_{0}\right]\left[d_{\text{h}}(t)/d_{\text{h,0}}\right]^{2}$. We follow typical convention \cite{Bear1972} in defining the hydraulic diameter $d_{\text{h}}(t)\sim A_{\text{pore}}(t)/P_{\text{pore}}(t)$, which we directly measure using confocal micrographs in a square region near the inlet of the medium approximately $400~\upmu$m $\times~400~\upmu$m across.  

\newpage\section*{Acknowledgements}
It is a pleasure to acknowledge Tapomoy Bhattacharjee for assistance with data analysis, Maziar Derakhshandeh for assistance with a preliminary version of the experimental platform, Bob Prud'homme for stimulating discussions, the developers of the Parti-Suite software package (\url{https://wpjohnsongroup.utah.edu/trajectoryCodes.html}) for providing this open-access resource, and the anonymous reviewers for constructive and insightful feedback. This work was supported by the Grand Challenges Initiative of the Princeton Environmental Institute, the Alfred Rheinstein Faculty Award, and a postdoctoral fellowship from the Princeton Center for Complex Materials (PCCM) to N.B. \\

\noindent \textbf{Author contributions:} N.B. and S.S.D. designed the experiments; N.B. performed experiments; N.B., J.S., and S.S.D. analyzed the data; N.B. and S.S.D. developed the theoretical model; S.S.D. designed and supervised the overall project. All authors discussed the results and wrote the manuscript. \\

\noindent \textbf{Competing interests:} The authors declare no competing interests. \\

\noindent \textbf{Data and materials availability:} All data are available in the manuscript or the supplementary materials and are available from the corresponding author on a reasonable request.\\

\newpage\section*{List of Supplementary Materials}
\noindent\textbf{Supplementary discussion.} Further discussion of particle forces, rescaling of deposition profiles, mechanistic particle trajectory simulations, and error analysis.

\noindent\textbf{Movie S1.} Extended deposition of amine-functionalized colloidal particles throughout a 3D porous medium.

\noindent\textbf{Movie S2.} Deposition of amine-functionalized colloidal particles within a single pore. 

\noindent\textbf{Movie S3.} Erosion of deposited amine-functionalized colloidal particles within a single pore. 

\noindent\textbf{Movie S4.} Localized deposition of amine-functionalized colloidal particles near the inlet of a 3D porous medium.

\noindent\textbf{Movie S5.} Extended deposition of carboxyl-functionalized colloidal particles throughout a 3D porous medium.

\noindent\textbf{Movie S6.} Deposition of carboxyl-functionalized colloidal particles within a single pore. 

\noindent\textbf{Movie S7.} Erosion of deposited carboxyl-functionalized colloidal particles within a single pore.

\noindent\textbf{Movie S8.} Localized deposition of carboxyl-functionalized colloidal particles near the inlet of a 3D porous medium.

\begin{figure}
    \centering
    \includegraphics[width=\textwidth]{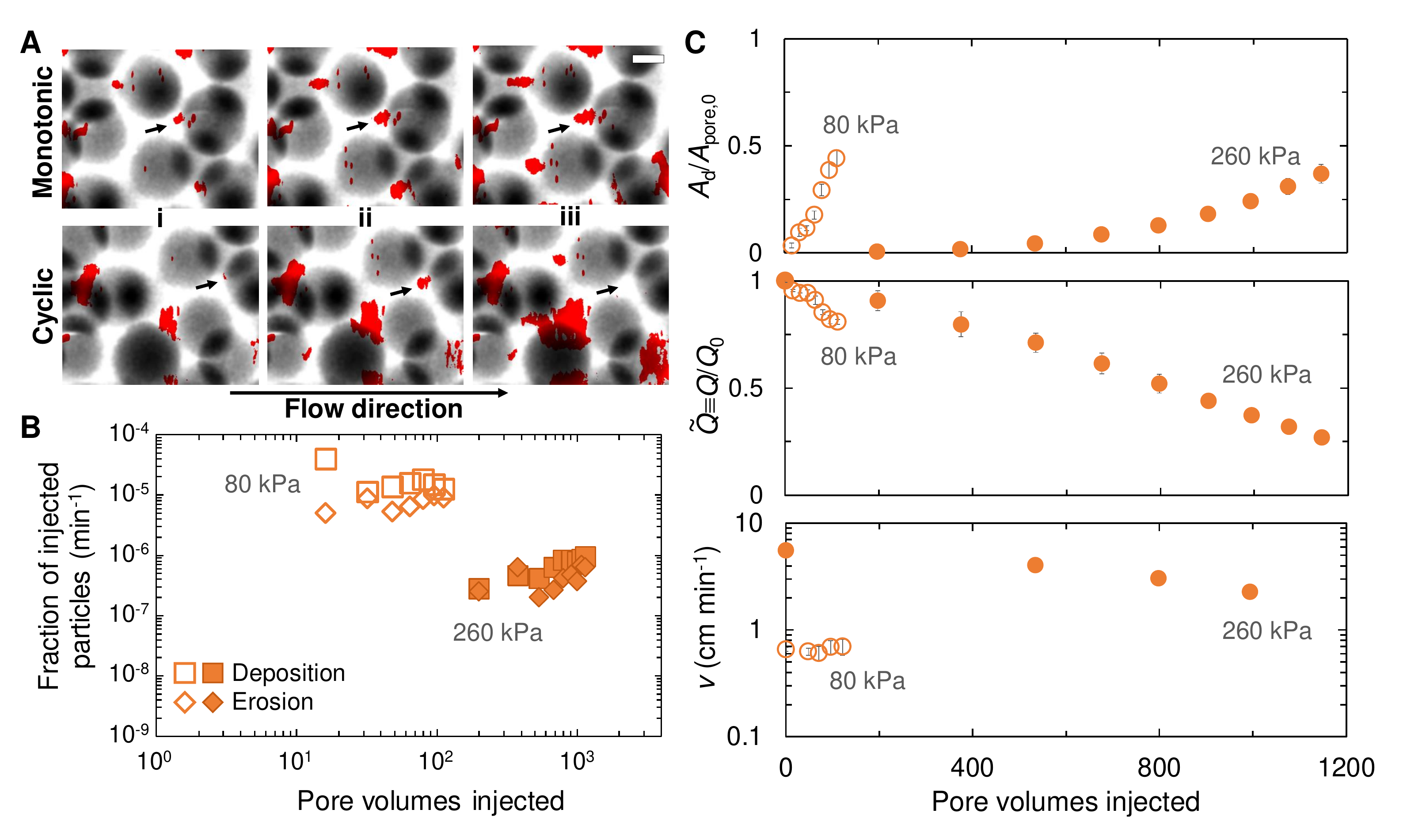}    
    \caption{\textbf{Deposition and erosion of positively-charged colloidal particles at the pore scale.} (A) Arrows in the upper and lower sequences of micrographs show monotonic deposition and cyclic deposition-erosion of particles upstream of a bead, respectively, at $260$ kPa. (i-iii) show micrographs after the injection of $675$, $903$, and $1074$ total pore volumes of the colloidal suspension. Black circles show cross-sections through the beads making up the porous media, white space shows pore space, and red shows deposited colloidal particles. Scale bar represents $20~\upmu$m. (B) Rates of particle deposition (squares) and erosion (diamonds), calculated by subtracting successive images, are shown as a function of total pore volumes injected for $260$ and $80$ kPa, indicated by the filled and open symbols, respectively. (C) Top, middle, and bottom panels show the fraction of the pore space area occupied by deposited particles $A_{\text{d}}/A_{\text{pore},0}$, the volumetric flow rate $Q$ normalized by its initial value ($1.5$ mL/h for $260$ kPa and $0.15$ mL/h for $80$ kPa), and the interstitial flow speed, respectively, as they vary with the total pore volumes injected. Experiments at $260$ and $80$ kPa are indicated by filled and open symbols, respectively. All data are for amine-functionalized polystyrene particles. Error bars in (C) reflect uncertainty arising from binarizing the micrographs or from variation in flow rate measurements, as detailed in the \textit{Supplementary Materials}. Error bars that are not observable in (C) are smaller than the symbol size.}
    \label{fig3}
\end{figure}

\begin{figure}
    \centering
    \includegraphics[width=\textwidth]{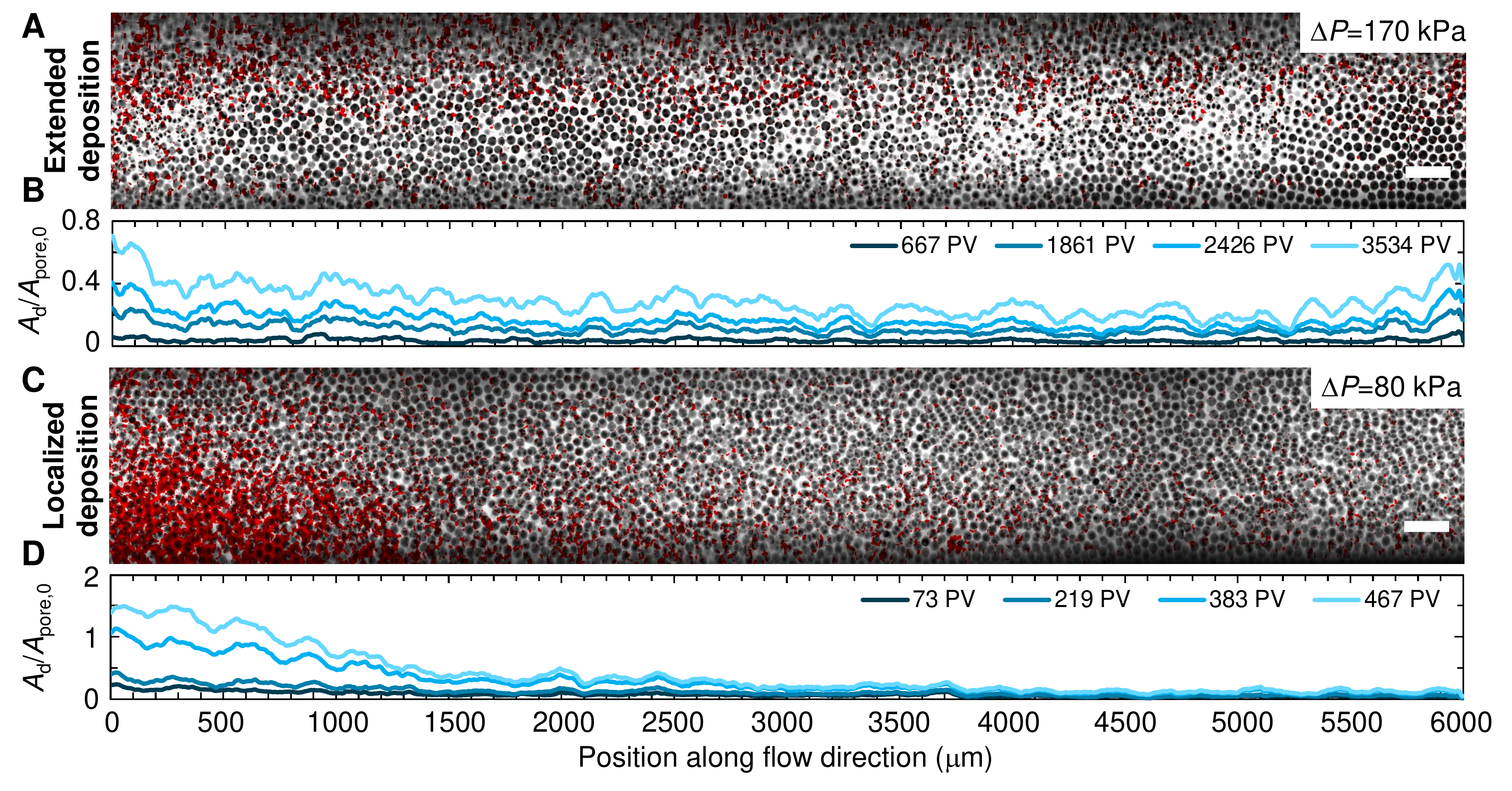}    
    \caption{\textbf{Macroscopic deposition profile of negatively-charged colloidal particles is tuned by injection pressure.} Confocal micrographs show extended (A) and localized (C) deposition profiles after $t = 5.6$ h and $16.7$ h for carboxyl-functionalized polystyrene particles injected at $170$ or $80$ kPa, respectively. Black circles show cross-sections through the beads making up the porous media, white space shows pore space, and red shows deposited colloidal particles. Corresponding line traces in (B) and (D) show the amount of deposition at each position along the flow direction, quantified by $A_{\text{d}}/A_{\text{pore},0}$ where $A_{\text{d}}$ is the laterally-averaged area occupied by deposited particles and $A_{\text{pore},0}$ is the pore space area before colloidal injection. Traces show $A_{\text{d}}$ with a running average over every $100~\upmu$m along the length of the medium to minimize noise; different traces are shown for images obtained every $3$ min in (B) and $2$ min in (D), which we represent by the total number of suspension pore volumes (PVs) injected. The traces exhibit some spatial fluctuations that likely reflect the influence of packing heterogeneities in the disordered media. In some cases, $A_{\text{d}}/A_{\text{pore},0}$ slightly exceeds one due to the non-zero thickness of the optical slice. For clarity, we only show the deposition profile for the first $6000~\upmu$m of the media, shorter than their full lengths. Flow direction is from left to right and the scale bars represent $200~\upmu$m.}
    \label{fig4}
\end{figure}

\begin{figure}
    \centering
    \includegraphics[width=\textwidth]{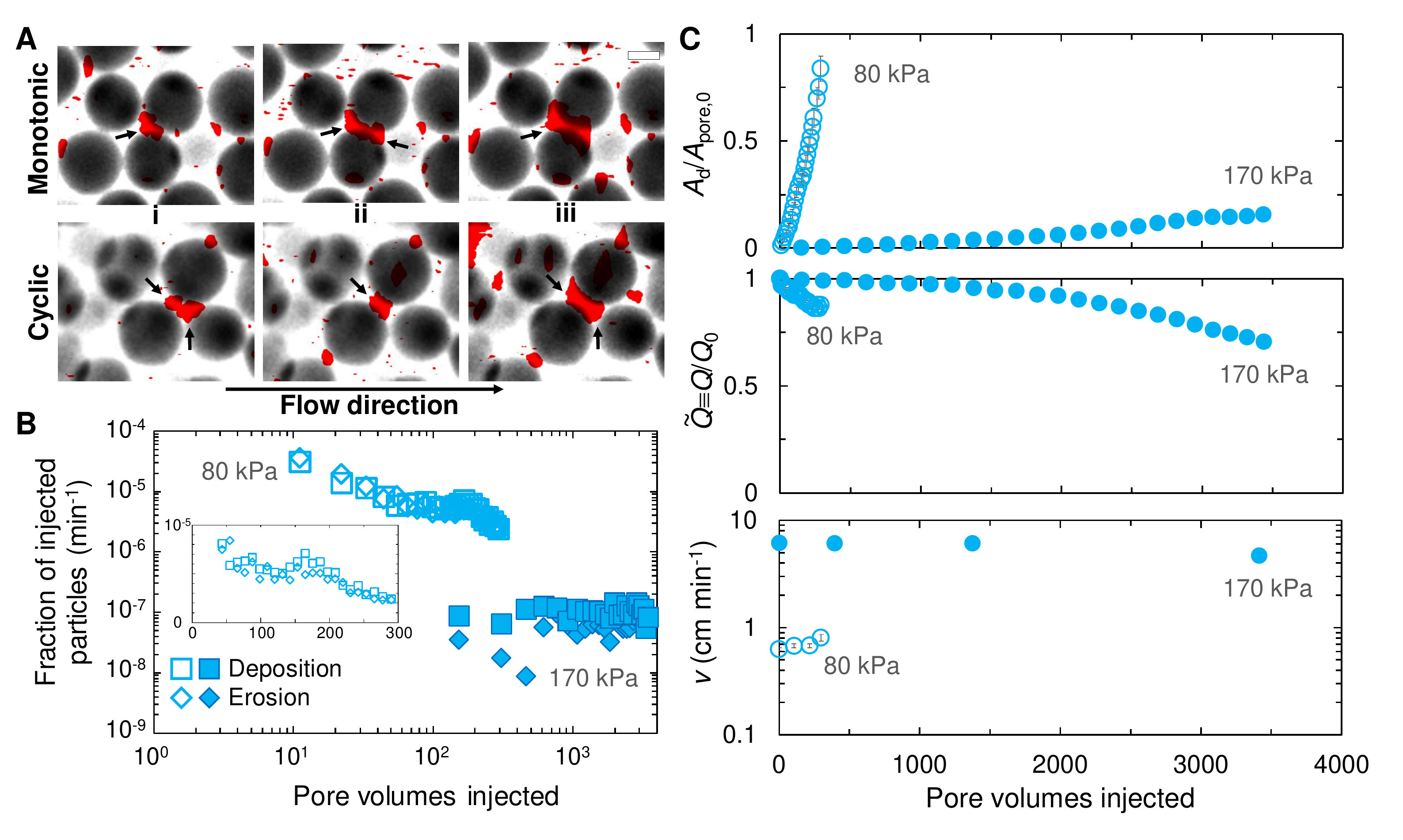}    
   \caption{\textbf{Deposition and erosion of negatively-charged colloidal particles at the pore scale.} (A) Arrows in the upper and lower sequences of micrographs show monotonic deposition and cyclic deposition-erosion of particles upstream of a bead, respectively, at $170$ kPa. (i-iii) show micrographs after the injection of $610$, $1376$, and $2603$ total pore volumes of colloidal suspension. Black circles show cross-sections through the beads making up the porous media, white space shows pore space, and red shows deposited colloidal particles. Scale bar represents $20~\upmu$m. (B) Rates of particle deposition (squares) and erosion (diamonds), calculated by subtracting successive images, are shown as a function of total pore volumes injected for $170$ and $80$ kPa, indicated by the filled and open symbols, respectively. (C) Top, middle, and bottom panels show the fraction of the pore space area occupied by deposited particles $A_{\text{d}}/A_{\text{pore},0}$, the volumetric flow rate $Q$ normalized by its initial value ($1.2$ mL/h for $170$ kPa and $0.16$ mL/h for $80$ kPa), and the interstitial flow speed, respectively, as they vary with the total pore volumes injected. Experiments at $170$ and $80$ kPa are indicated by filled and open symbols, respectively. All data are for carboxyl-functionalized polystyrene particles. Error bars in (C) reflect uncertainty arising from binarizing the micrographs or from variation in flow rate measurements, as detailed in the \textit{Supplementary Materials}. Error bars that are not observable in (C) are smaller than the symbol size.}
    \label{fig5}
\end{figure}

\begin{figure}
    \centering
    \includegraphics[width=\textwidth]{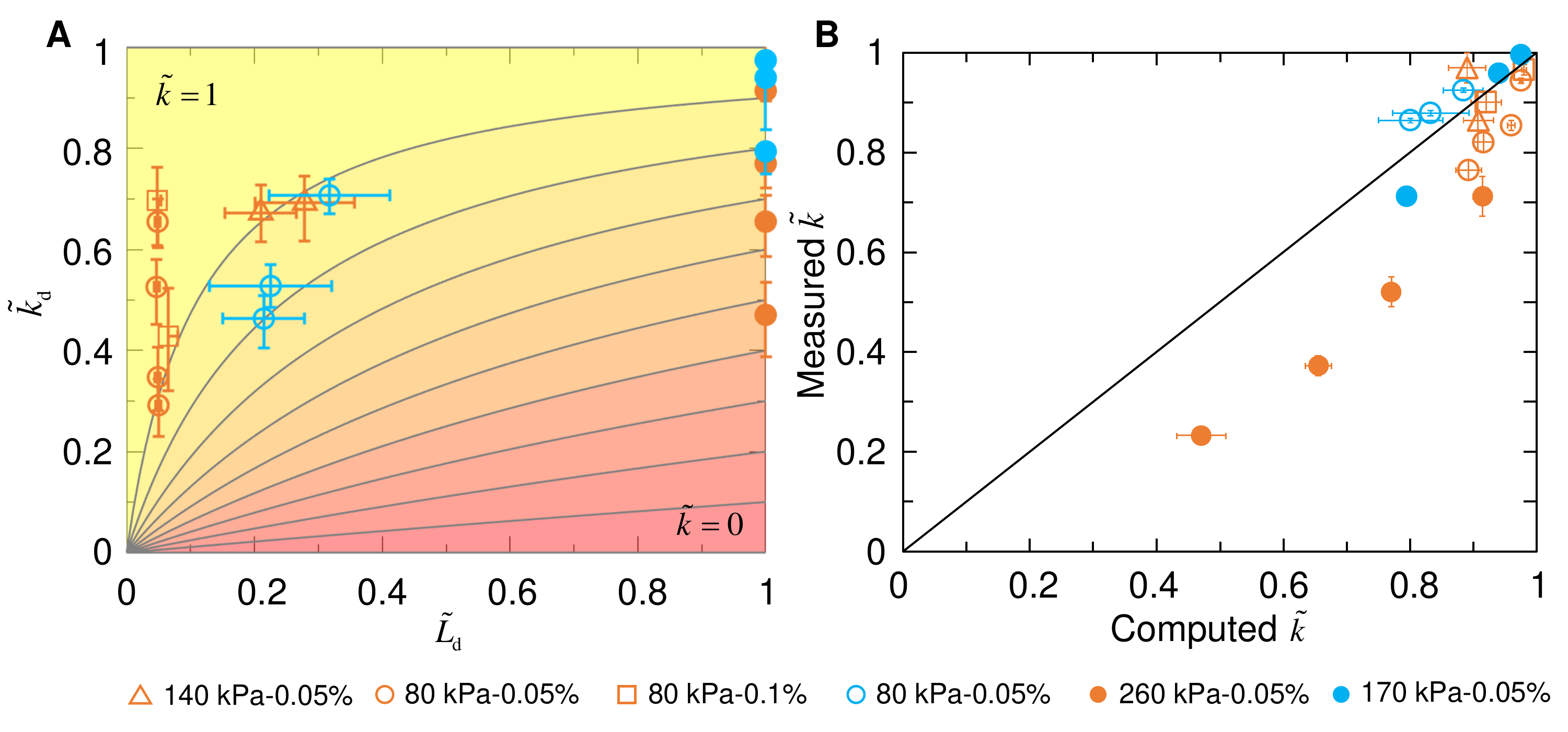}    
   \caption{\textbf{Pore-scale and macroscopic features of colloidal deposition control overall flow behavior.} (A) Color map showing the normalized overall permeability of the medium, $\tilde{k}\equiv k/k_0$, calculated using Eq. \ref{eq2} of the main text for different values of the normalized permeability and length of the particle-deposited region, $\tilde{k}_{\text{d}}\equiv k_{\text{d}}/k_0$ and $\tilde{L}_{\text{d}}\equiv L_{\text{d}}/L$, respectively. Yellow represents $\tilde{k}=1$ while pink represents $\tilde{k}=0$. (B) Measurements of the overall permeability of the medium, represented by the vertical axis, are in reasonable agreement with the prediction of Eq. \ref{eq2}, represented by the horizontal axis. Diagonal line shows a one-to-one relation. The deviation of the measurements from this relation as deposition progresses and $\tilde{k}$ decreases likely arises from the simplifying assumption that the deposited and pristine regions can be treated separately, with a homogeneous permeability in each; in reality, the amount of deposition varies spatially as well. Different points in both panels represent different experiments carried out for amine- or carboxyl-functionalized particles (orange or blue), at different imposed pressures, and using different particle volume fractions, as indicated by the legend underneath. Multiple points show different time points of each experiment. Error bars reflect uncertainty arising from binarizing the micrographs, defining a threshold for $L_{\text{d}}$, or from variation in flow rate measurements, as detailed in the \textit{Supplementary Materials}. }
    \label{fig6}
\end{figure}

\newpage
\setcounter{figure}{0}
\makeatletter 
\renewcommand{\thefigure}{S\@arabic\c@figure}
\makeatother

\section*{Supplementary Materials}\pagestyle{empty}
\noindent\textbf{DLVO calculations.} We use the DLVO framework to examine the interactions between different colloidal particles as well as between a colloidal particle and a glass bead. Specifically, the force $F_{\text{DLVO}}$ between two spherical bodies (1) and (2) having diameters $d_1$ and $d_2$, respectively, interacting through a medium (3) at a surface-to-surface distance of $h$ is given by:
\begin{equation*}
    \begin{aligned}
    F_{\text{DLVO}}(h)= \pi\frac{d_1 d_2}{d_1+d_2}\frac{\varepsilon_0\varepsilon_{\text{r}}\kappa\left[2\varPsi_1\varPsi _2-\left(\varPsi_1^2+\varPsi_2^2\right)e^{-\kappa h}\right]}{e^{+\kappa h}-e^{-\kappa h}}-\frac{H_{132}}{12h^2}\frac{d_1 d_2}{d_1+d_2},
    \label{DLVO}
    \end{aligned}
\end{equation*}
\noindent where the first and second terms on the right hand side represent electrostatic and van der Waals interactions, respectively; $\varepsilon_0$ is the vacuum permittivity; $\varepsilon_{\text{r}}$ is the dielectric constant of the fluid medium; $\kappa^{-1}$ is the Debye length of the fluid medium; $\varPsi_1$ and $\varPsi_2$ are the surface charges of bodies (1) and (2), respectively, which we approximate by their zeta potentials; and $H_{132}$ is the Hamaker constant representing the interaction between (1) and (2) through the medium (3). We use measured values of these parameters, summarized in Table S1, to estimate $F_{\text{DLVO}}$ for four different cases.

\begin{figure}
    \centering
    \includegraphics[width=\textwidth]{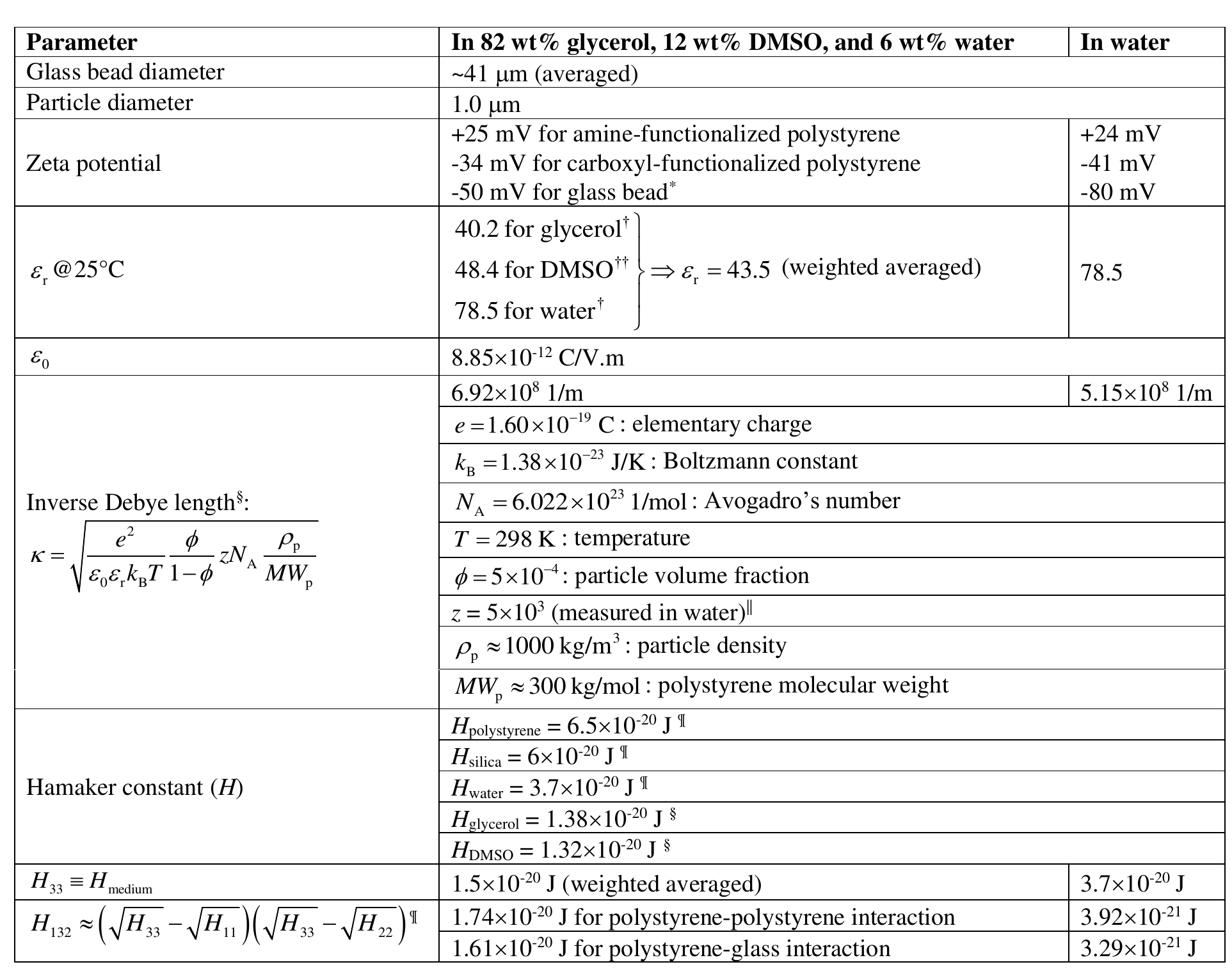}    
    \caption*{\textbf{Table S1: Parameters used for DLVO calculations.} Values indicated by the symbols $^{\ast}$, $^{\dagger}$, $^{\dagger\dagger}$, $^{||}$, $^{\mathparagraph}$, and $^{\mathsection}$ are obtained from \cite{Kosmulski2000}, \cite{Akerlof1932}, \cite{Puranik1992},\cite{Gogelein2012}, \cite{Gong2001}, and \cite{Israelachvili2011}, respectively.}
    \label{figtableS1}
\end{figure}

\begin{figure}
    \centering
    \includegraphics[width=\textwidth]{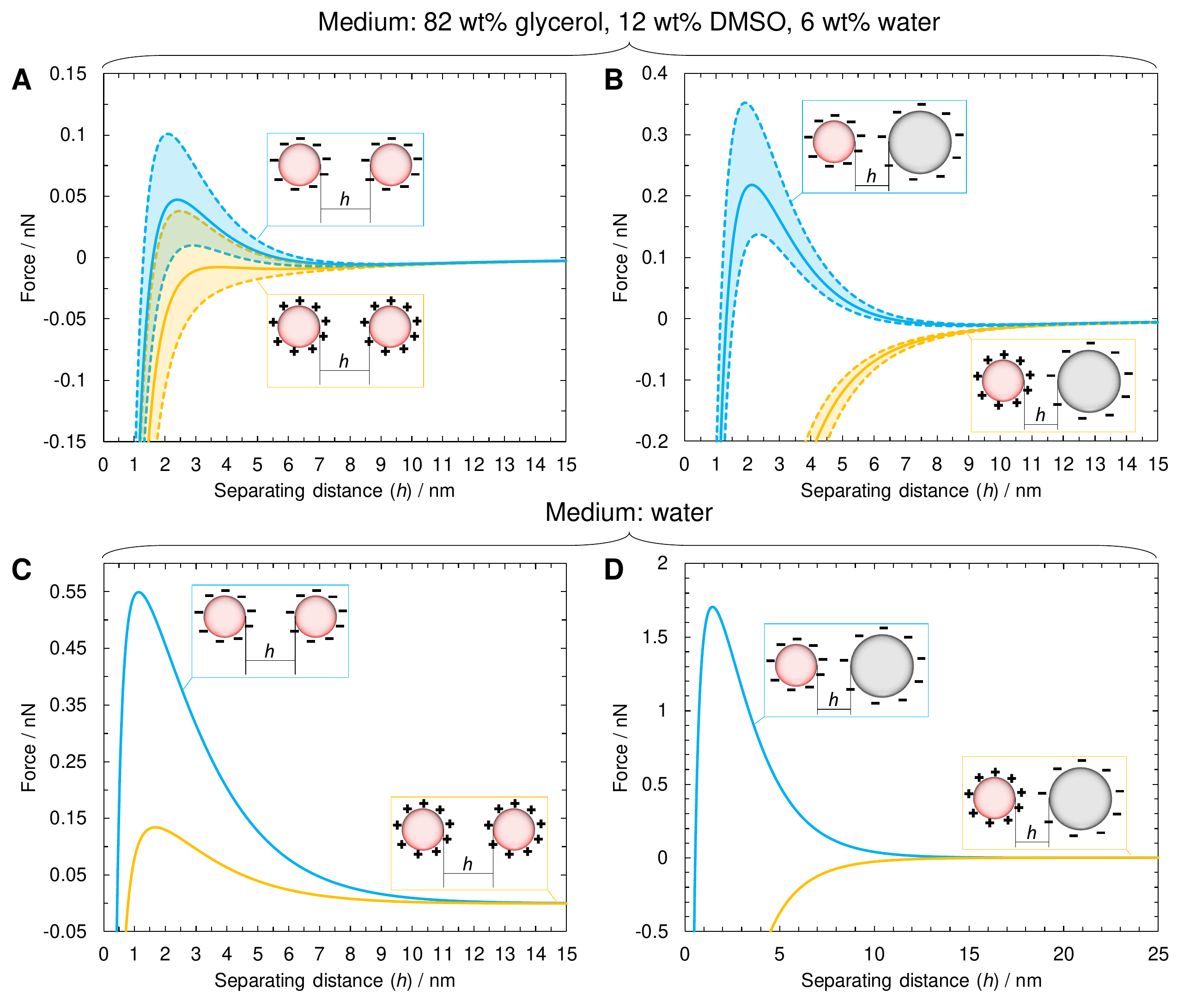}    
    \caption{\textbf{DLVO computations.} Total DLVO interactions (electrostatic and van der Waals forces) are computed for amine- or carboxyl-functionalized polystyrene particles interacting with each other (A) or with a glass bead (B) through the refractive index-matched liquid medium composed of 82 wt\% glycerol, 12 wt\% dimethyl sulfoxide (DMSO), and 6 wt\% water. (C) and (D) show the same results, but for a liquid medium of pure water. In all cases, like-charged bodies show some electrostatic repulsion when interacting with each other with van der Waals attraction at shorter separations, while interactions between oppositely-charged bodies are purely attractive. The shaded regions in (A) and (B) reflect the uncertainty in the DLVO curves arising from the experimental uncertainty in the measured zeta potentials.}
    \label{figS1}
\end{figure}

\textit{(A) Individual amine- or carboxyl-functionalized particles interacting with each other through the refractive index-matched fluid mixture}. The calculations indicate that interactions between amine-functionalized particles may have slight electrostatic repulsion at a separation of $\sim3$ nm, as shown by the yellow curves in Figure S1A, while those between carboxyl-functionalized particles have stronger electrostatic repulsion at a separation of $\sim2$ nm, as shown by the blue curves in Figure S1A, with slight attraction at larger separations in both cases. Further, as predicted by the stability criteria for a colloidal suspension that both $F_{\text{DLVO}}(h)=0$ and $dF_{\text{DLVO}}(h)/dh=0$ \cite{Lin2014}, the amine- and carboxyl-functionalized particle suspensions are expected to remain stable in suspension for zeta potential magnitudes larger than 27 mV, consistent with the measured zeta potential values and our observation that both suspensions remain stable for more than 6 days, much longer than the experimental duration.
 
  \textit{(B) Single amine- or carboxyl-functionalized polystyrene particle interacting with a glass bead through the refractive index-matched fluid mixture}. The calculations indicate that interactions between amine-functionalized particles and glass beads are purely attractive, as shown by the yellow curves in Figure S1B, while those between carboxyl-functionalized particles and glass beads have appreciable electrostatic repulsion at a separation of $\sim2$ nm, as shown by the blue curves in Figure S1B.

  \textit{(C) Individual amine- or carboxyl-functionalized polystyrene particles interacting with each other through pure water}. The calculations indicate that, similar to the case of the refractive index-matched fluid mixture, interactions between amine- and carboxyl-functionalized particles have appreciable electrostatic repulsion at a separation of $\sim1$ to $2$ nm, as shown by the yellow and blue curves in Figure S1C, respectively.

  \textit{(D) Single amine- or carboxyl-functionalized polystyrene particle interacting with a glass bead through pure water}. The calculations indicate that, similar to the case of the refractive index-matched fluid mixture, interactions between amine-functionalized particles and glass beads are purely attractive, as shown by the yellow curve in Figure S1D, while those between carboxyl-functionalized particles and glass beads have appreciable electrostatic repulsion at a separation of $\sim1$ to $2$ nm, as shown by the blue curve in Figure S1D.\\

\begin{figure}
    \centering
    \includegraphics[width=\textwidth]{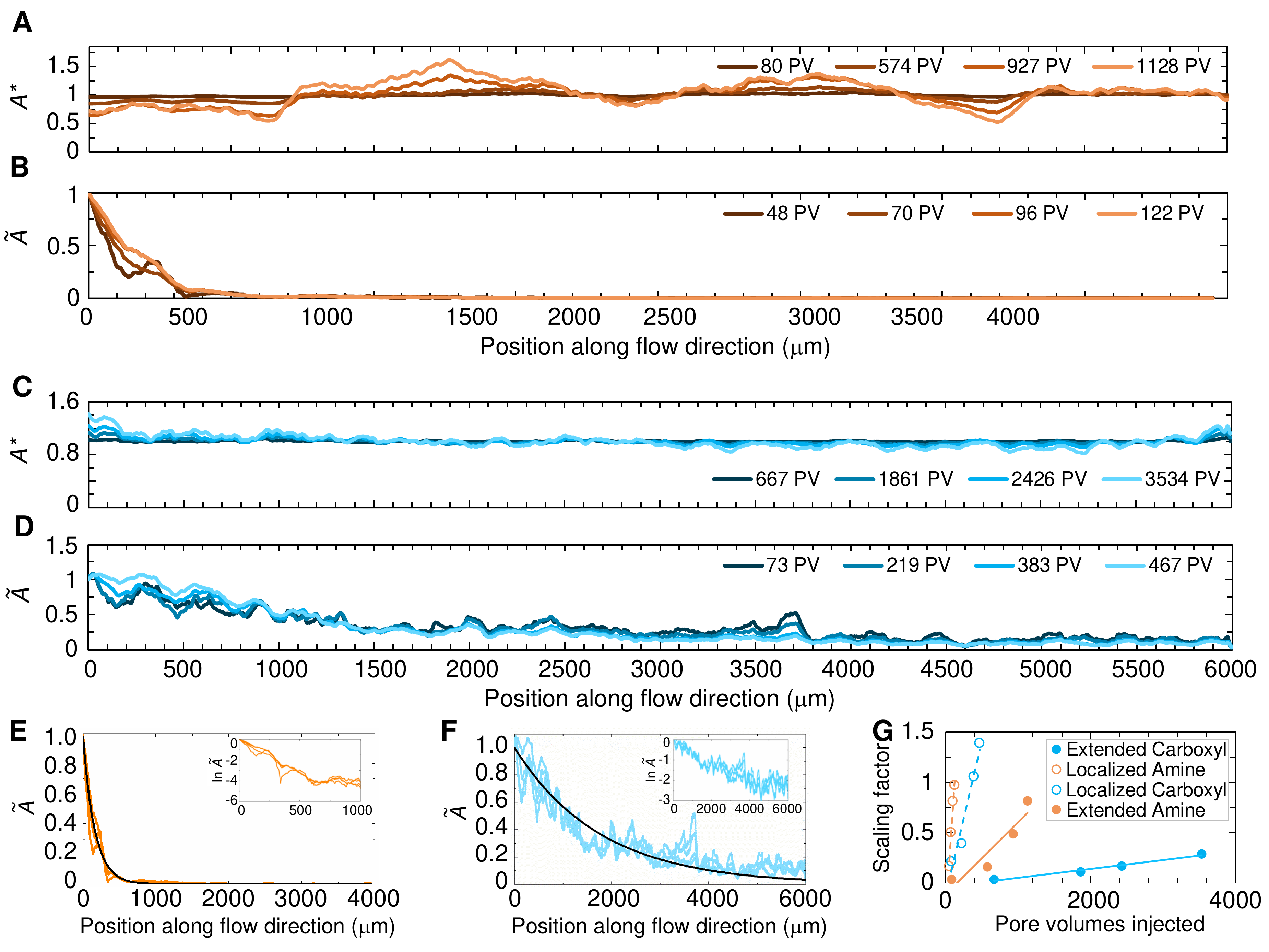}    
    \caption{\textbf{Rescaled deposition profiles.} The rescaled profiles showing the variation in the amount of deposition, $A^{\ast}$ or $\tilde{A}$, collapse for all the PVs tested. We define $A^{\ast}(x,t)\equiv A_{\text{d}}(x,t)/A_{\text{pore},0}+\big[1-\big<A_{\text{d}}(t)/A_{\text{pore},0}\big>_{x}\big]$, where $\big<A_{\text{d}}(t)/A_{\text{pore},0}\big>_{x}$ is $A_{\text{d}}(t)/A_{\text{pore},0}$ averaged over the entire medium at a given time $t$, and $\tilde{A}\equiv \big[A_{\text{d}}(x,t)/A_{\text{pore},0}\big]/\big[A_{\text{d}}(x=0,t)/A_{\text{pore},0}\big]$. Panels (A-B) correspond to panels (B) and (D) of the main text Figure 2, while panels (C-D) correspond to panels (B) and (D) of the main text Figure 4. As shown in (E-F), the localized deposition profiles appear to follow an exponential decay, indicated by the solid black lines and shown magnified in the insets. The fits in (E) and (F) are given by $\tilde{A}=e^{-6.27\times10^{-3}x}$ and $\tilde{A}=e^{-5.65\times10^{-4}x}$, respectively, where $x$ is the position along the flow direction in micrometers. The shift factors used to rescale the data in panels (A-D) scale linearly with the different PVs injected, as shown in panel (G). }
    \label{figS2}
\end{figure}


\noindent\textbf{Rescaled deposition profiles.} Previous work \cite{Gerber2018,Johnson2018,Johnson2020} indicates that for a filtration process that is consistent across time, $A_{\text{d}}/A_{\text{pore,0}}\sim\frac{pN_{i}}{N}e^{-px/d}$, where $p$ is the net deposition probability, $N_{i}$ is the number of injected particles over time $t$ and is therefore proportional to the PVs injected, $N\approx\phi_{0}w^{2}d/\left(\frac{1}{6}\pi d_{\text{p}}^{3}\right)$, and $x$ represents the position along the flow direction. Our pore-scale imaging of colloidal deposition and erosion suggest that the net deposition is indeed consistent over time; we therefore expect similar scaling to hold in our experiments. We test this expectation for the deposition profiles shown in Figures 2B,D and 4B,D. Specifically, for the case of extended deposition, we define the rescaled amount of deposition as $A^{\ast}(x,t)\equiv A_{\text{d}}(x,t)/A_{\text{pore},0}+\big[1-\big<A_{\text{d}}(t)/A_{\text{pore},0}\big>_{x}\big]$, where $\big<A_{\text{d}}(t)/A_{\text{pore},0}\big>_{x}$ is $A_{\text{d}}(t)/A_{\text{pore},0}$ averaged over the entire medium at a given time $t$; for the case of localized deposition, we define the rescaled amount of deposition as $\tilde{A}\equiv \big[A_{\text{d}}(x,t)/A_{\text{pore},0}\big]/\big[A_{\text{d}}(x=0,t)/A_{\text{pore},0}\big]$. With the exception of slight fluctuations likely reflecting the influence of packing inhomogeneities, the rescaled deposition profiles indeed appear to collapse onto each other, as shown in Figures S2A-D. Moreover, the localized deposition profiles appear to exhibit the predicted exponential decay, indicated by the dark lines in Figures S2E-F; comparing these exponential fits to the prediction that deposition varies as $\sim e^{-px/d}$ yields $p=0.25$ and $p=0.023$ for amine- and carboxyl-functionalized particles, respectively. Finally, the scaling factors used to rescale the datasets \textit{i.e.}, $\big<A_{\text{d}}(t)/A_{\text{pore},0}\big>_{x}$ and $A_{\text{d}}(x=0,t)/A_{\text{pore},0}$ for the cases of extended and localized deposition, respectively, increase linearly with the number of suspension pore volumes injected, as predicted. Together, these data indicate that the process by which particles are deposited is consistent over time \cite{Gerber2018,Johnson2018,Johnson2020}.  \\

\noindent\textbf{Mechanistic particle trajectory simulations using Parti-Suite.} We use the Parti-Suite framework \cite{PartiSuite} to examine the interactions between single particles or clusters of particles and the glass beads making up the solid matrix. In this framework, Lagrangian trajectories that explicitly incorporate the forces and torques on particles \cite{VanNess2019} are simulated in a Happel sphere-in-cell model of the pore space surrounding an individual bead \cite{Molnar2015}. Previous work has established the ability of this framework to quantitatively model particle deposition and erosion from surfaces under the influence of imposed flow \cite{VanNess2019}. Specifically, the particles are initialized at random points at the entrance of the pore upstream of the bead. We choose the flow velocity to match that corresponding to that of our experiments. Prior to particle-bead contact, particle velocity is computed from the action of hydrodynamic forces exerted by the imposed flow, computed using the parameter values given in Table S2. Upon contact with the bead surface, particle movement is dictated by the balance between hydrodynamic and attachment torques \cite{VanNess2019}, computed using the parameter values given in Tables S1 and S2, respectively. The trajectories of multi-particle clusters are treated as larger $4~\upmu$m-diameter particles simulated in the same way, but using physico-chemical parameters that describe the relevant colloid-colloid interactions. We use these simulations to assess deposition and erosion in four different cases, described below.

\textit{(1) Individual or clusters of amine-functionalized polystyrene particles in the refractive index-matched fluid mixture}. Out of ten different simulated trajectories of $1~\upmu$m amine-functionalized polystyrene particles approaching a $\sim40~\upmu$m glass bead at a high imposed pressure ($\Delta P=260$ kPa), 60\% of particles do not attach to the bead surface, as shown by the red lines in Figure S3Ai. The other 40\% are irreversibly deposited on the upstream ($Z>0$) bead surface, as shown by the blue lines in Figure S3Ai, consistent with our experimental findings shown in Figure \ref{fig3}A and Movies S2-S3. Once a particle deposits onto the bead, the hydrodynamic torques are not large enough to detach it from the bead surface, as exemplified by the 3D trajectories of particles that discontinue due to deposition in Figure S3Aii. However, for the case of particle clusters interacting with a bead coated with a monolayer of particles at a high imposed pressure, the clusters deposit on the upstream ($Z>0$) bead surface, but are subsequently eroded away, as shown in Figure S3B. Hence, the particle trajectory simulations reproduce the continual deposition and erosion of particles observed at high pressures in our experiments. 

The case of low imposed pressure ($\Delta P=80$ kPa) is dramatically different. When ten amine-functionalized polystyrene particles are injected at a low imposed pressure, only 20\% of particles do not attach to the bead surface, while 80\% are irreversibly deposited on the upstream ($Z>0$) bead surface, as shown in Figure S3C. Thus, the relative influence of erosion is suppressed at lower pressures, consistent with our experimental results. Erosion is also suppressed at lower pressures for particle clusters where all clusters deposit on the upstream ($Z>0$) bead surface, roll over the bead, but remain attached, as shown in Figure S3D. Hence, the particle trajectory simulations reproduce the suppressed relative influence of erosion observed at low pressures in the experiments.

\textit{(2) Individual or clusters of carboxyl-functionalized polystyrene particles in the refractive index-matched fluid mixture}. Out of ten different simulated trajectories of $1~\upmu$m carboxyl-functionalized polystyrene particles approaching a $\sim40~\upmu$m glass bead at a high imposed pressure ($\Delta P=170$ kPa), all particles approach, but do not stick to, the bead surface, as shown in Figure S4A. This behavior is due to the electrostatic repulsion between particles and the bead surface, and is consistent with the experimental observation that instead of depositing on the upstream bead surface, carboxyl-functionalized particles are strained in the tight pore throats between beads, as shown in Figure \ref{fig5}A and Movies S6-S7. Similarly, for the case of particle clusters interacting with a bead coated with a monolayer of particles at a high imposed pressure, the clusters deposit on the upstream ($Z>0$) bead surface and are subsequently eroded away, as shown in Figure S4B. 

At a low imposed pressure ($\Delta P=80$ kPa), all ten particles again approach the bead surface, but do not stick to it, as shown in Figure S4C---again reflecting the electrostatic repulsion between particles and the bead surface. However, particle clusters deposit on the upstream ($Z>0$) bead surface, roll over the bead, but remain attached, as shown in Figure S4D. Therefore, the particle trajectory simulations reproduce the suppressed relative influence of erosion observed at low pressures in the experiments.

\textit{(3) Individual or clusters of amine-functionalized polystyrene particles in pure water}. We find similar behaviors to those obtained using the refractive index-matched fluid mixture in case (1), but at different values of the imposed fluid pressure, reflecting the influence of the different physico-chemical properties of the solvents; the flow velocity is larger by a factor of $50$ for both the high and low pressure cases. Similar to the simulations performed for particles in the refractive index-matched fluid mixture, we observe appreciable deposition and erosion of individual particles and of particle clusters at a high imposed pressure, as shown in Figures S5A-B. By contrast, at a lower imposed pressure, erosion is suppressed in both cases, as shown in Figures S5C-D. Thus, the particle trajectory simulations indicate that the experimentally-observed continual deposition and erosion at high pressures, and the suppressed relative influence of erosion at low pressures, also manifest in pure water-based colloids. 

\textit{(4) Individual or clusters of carboxyl-functionalized polystyrene particles in pure water}. We again find similar behaviors to those obtained using the refractive index-matched fluid mixture in case (2), but at different values of the imposed fluid pressure: for simulations in water, we increase the flow velocity by a factor of $50$ for both the high and low pressure cases. Similar to the simulations performed for particles in the refractive index-matched fluid mixture, we observe that individual particles or particle clusters approach the bead surface, but do not stick to it, as shown in Figures S6A and B---consistent with our pore-scale observation that these particles are strained in the tight pore throats between beads instead. At a lower imposed pressure, the individual particles approach the bead surface, but do not stick to it, as shown in Figure S6C; however, particle clusters deposit, roll, and remain attached to the bead surface, as shown in Figure S6D---just as in the simulations for the refractive index-matched fluid mixture. Hence, the particle trajectory simulations again indicate that our experimentally-observed continual deposition and erosion at high pressures, and the suppressed relative influence of erosion at low pressures, also manifest in pure water-based colloids.

\begin{figure}
    \centering
    \includegraphics[height=16.2cm]{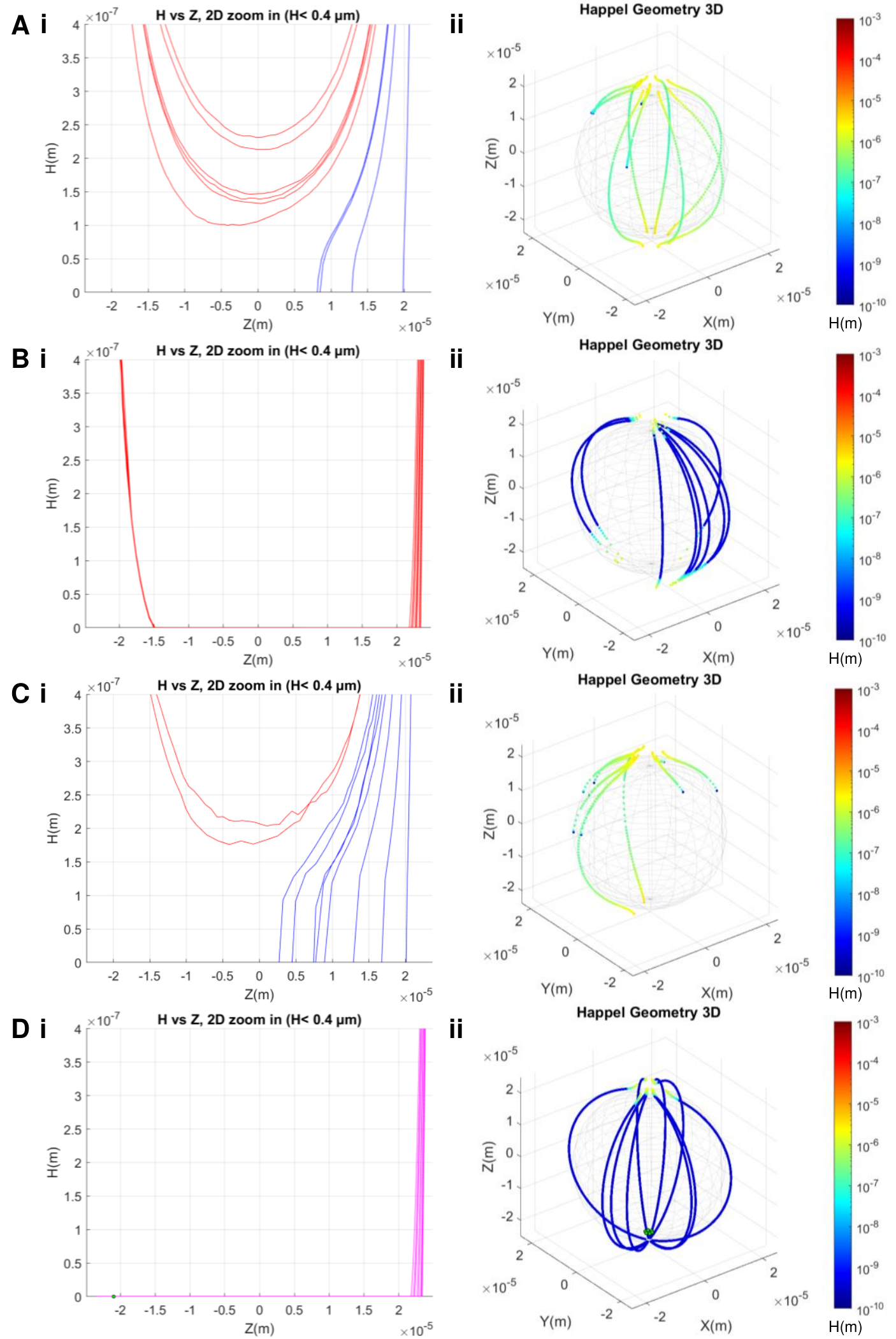}    
    \caption{\textbf{Simulations of amine-functionalized particles show that the relative influence of erosion is larger at larger imposed pressure.} (A) Ten different simulated trajectories of individual particles approaching a glass bead at a high imposed pressure. $H$ is the particle-bead separation distance and $X$, $Y$, and $Z$ are spatial coordinates with the origin $X=Y=Z=0$ located at the center of the bead and flow imposed along the $Z<0$ direction. Panel (i) shows the different trajectories in 2D; red and blue curves show particles that do not or do stick to the bead surface, respectively. Panel (ii) shows the different trajectories in 3D. (B) is the same as (A), but showing the trajectories of particle clusters instead. In this case, all particle clusters deposit on the upstream ($Z>0$) bead surface but are subsequently eroded away. (C) is the same as (A), but for the case of low imposed pressure. (D) is the same as (C), but showing the trajectories of particle clusters instead. In this case, all particle clusters deposit on the upstream ($Z>0$) bead surface, roll over the bead surface, but remain attached.  }
    \label{figS3}
\end{figure}

\begin{figure}
    \centering
    \includegraphics[height=16.2cm]{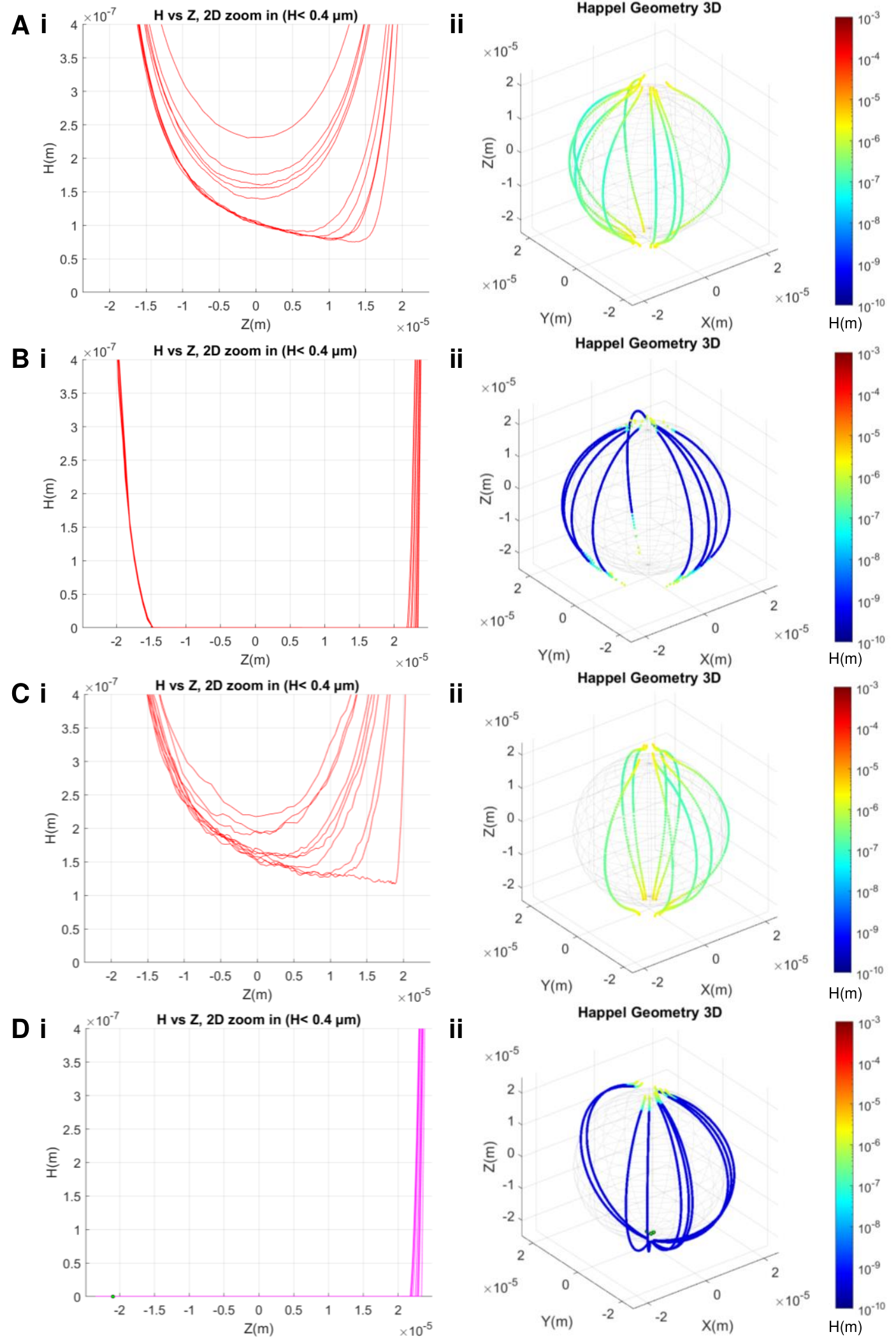}    
    \caption{\textbf{Simulations of carboxyl-functionalized particles show that the relative influence of erosion is larger at larger imposed pressure.} (A) Ten different simulated trajectories of individual particles approaching a glass bead, at a high imposed pressure. $H$ is the particle-bead separation distance and $X$, $Y$, and $Z$ are spatial coordinates with the origin $X=Y=Z=0$ located at the center of the bead and flow imposed along the $Z<0$ direction. Panel (i) shows the different trajectories in 2D; red curves show that all particles approach, but do not stick to, the bead surface. Panel (ii) shows the different trajectories in 3D. (B) is the same as (A), but showing the trajectories of particle clusters instead. In this case, all particle clusters deposit on the upstream ($Z>0$) bead surface but are subsequently eroded away. (C) is the same as (A), but for the case of low imposed pressure. (D) is the same as (C), but showing the trajectories of particle clusters instead. In this case, all particle clusters deposit on the upstream ($Z>0$) bead surface, roll over the bead surface, but remain attached.  }
    \label{figS4}
\end{figure}

\begin{figure}
    \centering
    \includegraphics[height=16.2cm]{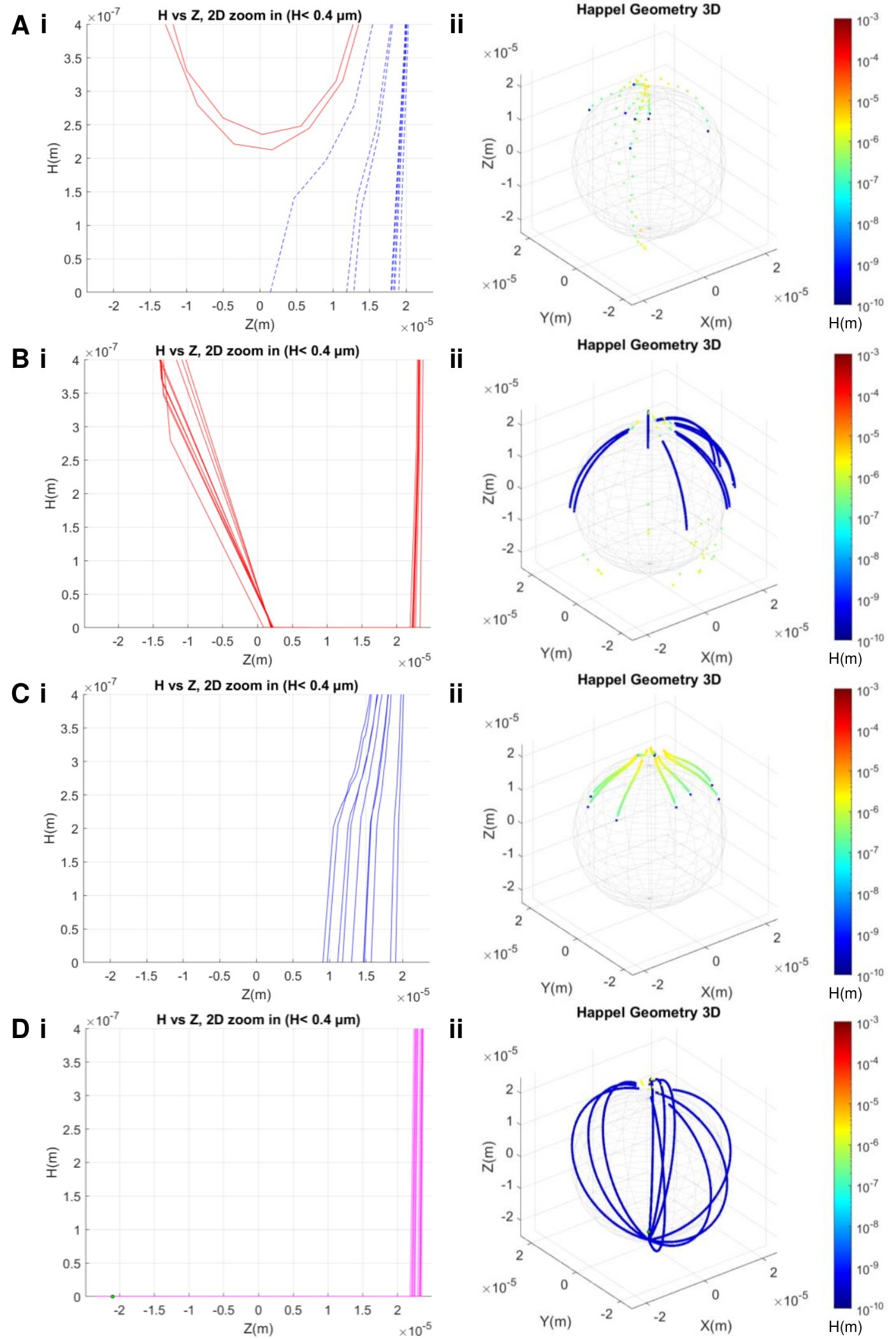}    
    \caption{\textbf{Simulations of amine-functionalized particles in water show similar behavior to the refractive index-matched mixture.} (A) Ten different simulated trajectories of individual particles approaching a glass bead at a high imposed pressure. $H$ is the particle-bead separation distance and $X$, $Y$, and $Z$ are spatial coordinates with the origin $X=Y=Z=0$ located at the center of the bead and flow imposed along the $Z<0$ direction. Panel (i) shows the different trajectories in 2D; red and blue curves show particles that do not or do stick to the bead surface, respectively. Panel (ii) shows the different trajectories in 3D. (B) is the same as (A), but showing the trajectories of particle clusters instead. In this case, all particle clusters deposit on the upstream ($Z>0$) bead surface but are subsequently eroded away. (C) is the same as (A), but for the case of low imposed pressure. All particles are irreversibly deposited on the upstream ($Z>0$) bead surface. (D) is the same as (C), but showing the trajectories of particle clusters instead. In this case, all particle clusters deposit on the upstream ($Z>0$) bead surface, roll over the bead surface, but remain attached.}
    \label{figS5}
\end{figure}

\begin{figure}
    \centering
    \includegraphics[height=16.2cm]{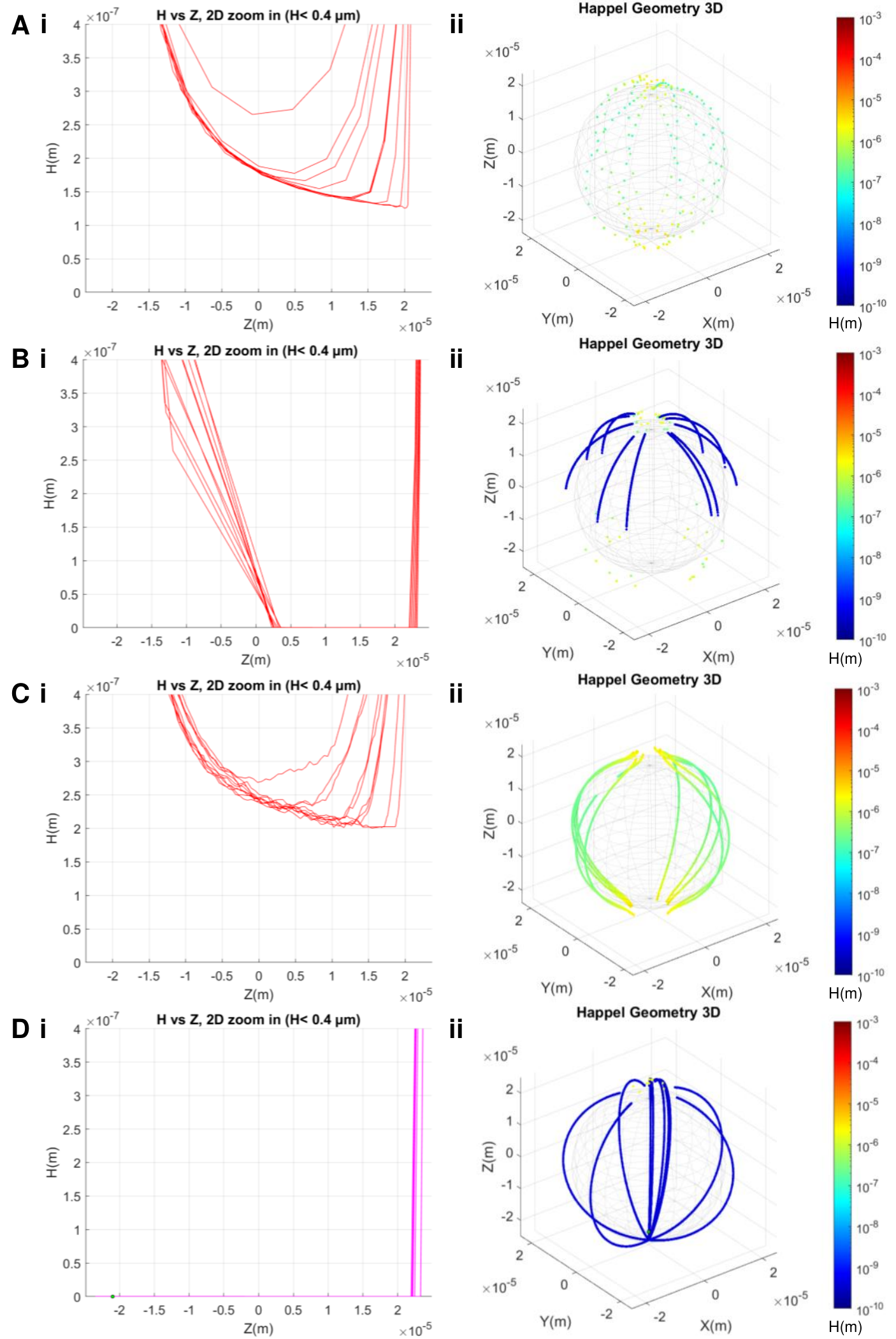}    
    \caption{\textbf{Simulations of carboxyl-functionalized particles in water show similar behavior to the refractive index-matched mixture.} (A) Ten different simulated trajectories of individual particles approaching a glass bead, at a high imposed pressure. $H$ is the particle-bead separation distance and $X$, $Y$, and $Z$ are spatial coordinates with the origin $X=Y=Z=0$ located at the center of the bead and flow imposed along the $Z<0$ direction. Panel (i) shows the different trajectories in 2D; red curves show that all particles approach, but do not stick to, the bead surface. Panel (ii) shows the different trajectories in 3D. (B) is the same as (A), but showing the trajectories of particle clusters instead. In this case, all particle clusters deposit on the upstream ($Z>0$) bead surface but are subsequently eroded away. (C) is the same as (A), but for the case of low imposed pressure. (D) is the same as (C), but showing the trajectories of particle clusters instead. In this case, all particle clusters deposit on the upstream ($Z>0$) bead surface, roll over the bead surface, but remain attached.}
    \label{figS6}
\end{figure}

\begin{figure}
    \centering
    \includegraphics[width=\textwidth]{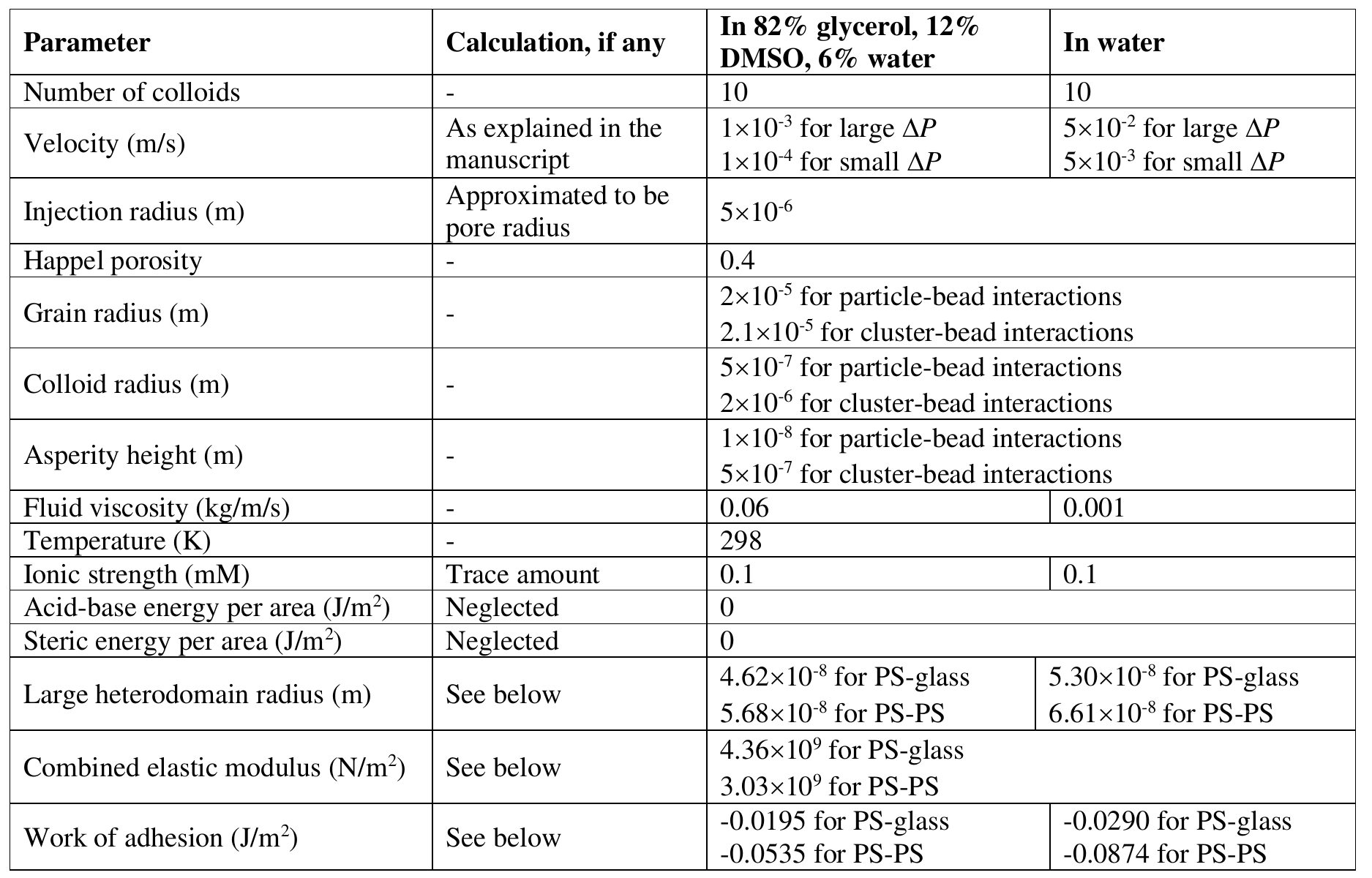}    
    \caption*{\textbf{Table S2: Additional parameters used as inputs in Parti-Suite particle trajectory simulations.} Work of adhesion ($W$) for body (1) interacting with body (2) through medium (3) is computed from \cite{VanNess2019}: $W=2\bigg[\sqrt{\gamma_1^{LW}\gamma_3^{LW}}+\sqrt{\gamma_2^{LW}\gamma_3^{LW}}-\sqrt{\gamma_1^{LW}\gamma_2^{LW}}-\gamma_3^{LW}+\sqrt{\gamma_1^{+}}\left(\sqrt{\gamma_1^{-}}+\sqrt{\gamma_2^{-}}-\sqrt{\gamma_3^{-}}\right)+$
    $\sqrt{\gamma_3^{-}}\left(\sqrt{\gamma_1^{+}}+\sqrt{\gamma_2^{+}}-\sqrt{\gamma_3^{+}}\right)-\sqrt{\gamma_1^{+}\gamma_2^{-}}-\sqrt{\gamma_1^{-}\gamma_2^{+}}\bigg]$, where $\gamma^{LW}$, $\gamma^{+}$, and $\gamma^{-}$ are 42 mJ/m$^{2}$, 0, and 1.1 mJ/m$^{2}$, respectively, for polystyrene (PS) \cite{VanNess2019}; 27.3 mJ/m$^{2}$, 0.1 mJ/m$^{2}$, and 37.5 mJ/m$^{2}$, respectively, for glass \cite{Hamadi2009}; 33.6 mJ/m$^{2}$, 8.41 mJ/m$^{2}$, and 31.16 mJ/m$^{2}$, respectively, for glycerol \cite{Fernandez2015}; and 21.8 mJ/m$^{2}$, 25.5 mJ/m$^{2}$, and 25.5 mJ/m$^{2}$, respectively, for water \cite{VanNess2019}. Combined elastic modulus ($K$) is computed from $K=\frac{4}{3\left[(1-v_1^2)/E_1+(1-v_2^2)/E_2\right]}$. For PS-glass system, $K = 4.36\times10^9$ N/m$^{2}$ \cite{VanNess2019}. For PS, $v_1$ (Poisson’s ratio) and $E_1$ (Young’s modulus) are 0.348 and 4 GPa \cite{Mott2008}, respectively, and thus, $K = 3.03\times10^9$ N/m$^{2}$ for for PS-PS. Heterodomains are nanoscale surface heterogeneity wherein net repulsion is reversed to net attraction when a nanoscale heterodomain occupies a critical fraction of the zone of colloid-bead interaction (ZOI) \cite{VanNess2019}. Heterodomain radius ($R_{\text{ZOI}}$) is calculated from $R_{\text{ZOI}}=\sqrt{a_{\text{cont}}^2+2\kappa^{-1}\left(r_{\text{p}}^2-a_{\text{cont}}^2\right)}$, where $\kappa^{-1}$ and $r_{\text{p}}$ are the Debye length and particle radius, respectively, and $a_{\text{cont}}=\left(\frac{6\pi Wr_{\text{p}}^2}{K}\right)^{1/3}$.}
    \label{figtableS2}
\end{figure}

\newpage\noindent\textbf{Error analysis.} The deposition length $\tilde{L}_{\text{d}}(t)$ is determined as the distance from the porous medium entrance beyond which $A_{\text{d}}/A_{\text{pore,0}}$ is smaller than a threshold value of $15\%$. To determine the uncertainty in $\tilde{L}_{\text{d}}(t)$, we consider two effects: (i) changes in this threshold value, and (ii) changes in the binarization of the confocal micrographs used to determine $A_{\text{d}}/A_{\text{pore,0}}$. To assess the influence of (i), we also determine $\tilde{L}_{\text{d}}(t)$ using threshold values of $10\%$ and $20\%$, shown by the columns in Table S3. To assess the influence of (ii), we also determine $\tilde{L}_{\text{d}}(t)$ with changes in the threshold value used to binarize the confocal micrographs by $\pm20\%$, shown by the rows in Table S3. Overall, we find that the threshold value of $A_{\text{d}}/A_{\text{pore,0}}$ has comparable or larger effects on $\tilde{L}_{\text{d}}(t)$ than those from binarization, and therefore, we use the variation arising from (i) to determine the uncertainty in $\tilde{L}_{\text{d}}(t)$, $\delta\tilde{L}_{\text{d}}(t)$. These values are represented by the horizontal error bars in Figure 6A. To estimate the uncertainty in $A_{\text{d}}/A_{\text{pore,0}}$, we vary the threshold value used to binarize the confocal micrographs by $\pm20\%$, shown by the error bars in the top panels of Figures 3C and 5C. To estimate the uncertainty in $Q(t)$, $\delta Q(t)$, we calculate the standard deviation of four measurements of $Q(t)$ at four successive times separated by 5 minutes preceding and up to time $t$, shown by the error bars in the middle panels of Figures 3C and 5C; this estimate also yields an estimate for the uncertainty in the measured $\tilde{k}(t)\equiv\tilde{Q}(t)$, shown by the vertical error bars in Figure 6B. These values also enable us to estimate the uncertainty in $v(t)\equiv Q(t)/\left[\phi(t)A\right]$ as $\delta v(t)/v(t)=\sqrt{\left[\delta Q(t)/Q(t)\right]^2+\left[\delta\phi(t)/\phi(t)\right]^2}$, where $\delta\phi(t)$ is estimated by varying the threshold value used to binarize the confocal micrographs by $\pm20\%$; the resultant uncertainty in $v$ is shown by the error bars in the bottom panels of Figures 3C and 5C. We also estimate the uncertainty in $\tilde{k}_{\text{d}}(t)$, $\delta\tilde{k}_{\text{d}}(t)$, by varying the threshold value used to binarize the confocal micrographs by $\pm20\%$ and determining the resultant changes in $\tilde{k}_{\text{d}}(t)$ using the protocol for calculating the permeability given in the \textit{Materials and Methods}, shown by the vertical error bars in Figure 6A. Finally, we estimate the uncertainty in the $\tilde{k}(t)$ computed from Eq. \ref{eq2} as
\begin{equation*}
    \begin{aligned}
    \delta\tilde{k}(t)=\frac{\tilde{L}_{\text{d}}(t)\left(1/\tilde{k}_{\text{d}}(t)-1\right)}{\left[1+\tilde{L}_{\text{d}}(t)\left(1/\tilde{k}_{\text{d}}(t)-1\right)\right]^2}\sqrt{\left[\delta\tilde{L}_{\text{d}}(t)/\tilde{L}_{\text{d}}(t)\right]^{2}+\left[\delta\tilde{k}_{\text{d}}(t)/\tilde{k}_{\text{d}}(t)\right]^{2}},
    \end{aligned}
\end{equation*}
\noindent shown by the horizontal error bars in Figure 6B.

\begin{figure}
    \centering
    \includegraphics[width=16cm]{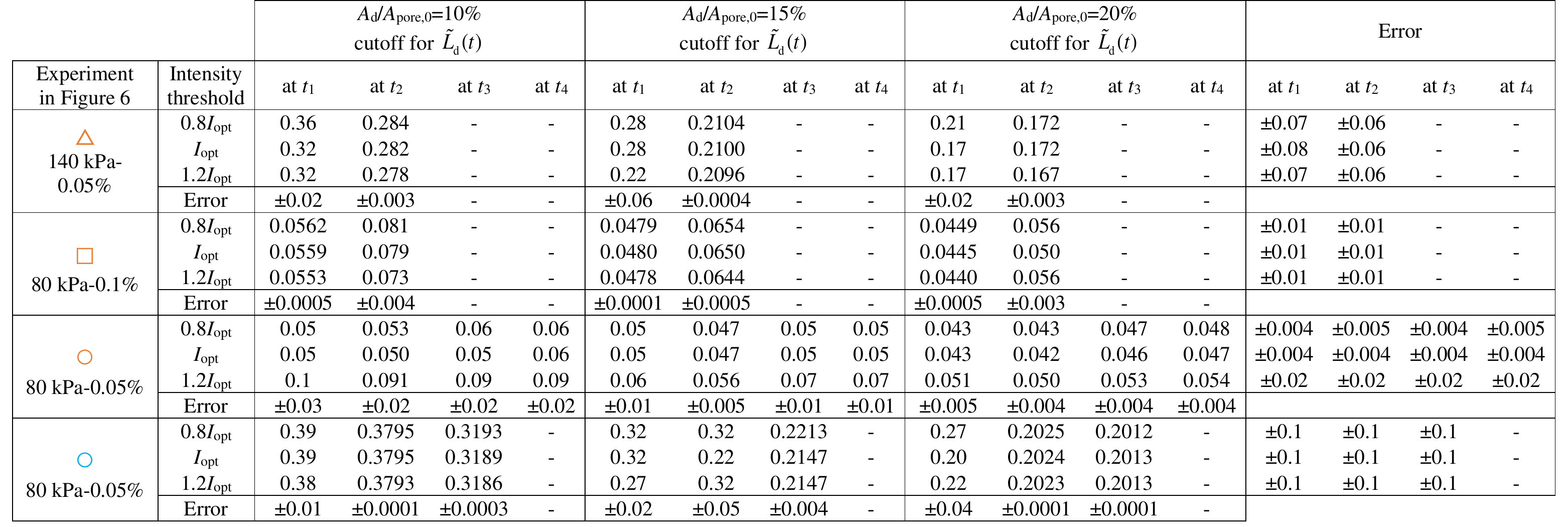}    
    \caption*{\textbf{Table S3: Sensitivity analysis of the deposition length.} $\tilde{L}_{\text{d}}(t)$ is determined as the distance from the porous medium entrance beyond which $A_{\text{d}}/A_{\text{pore,0}}$ is smaller than a threshold value of $15\%$. To determine the uncertainty in $\tilde{L}_{\text{d}}(t)$, we assess the effect of varying this threshold value between $10$ and $20\%$, as shown by the columns, or varying the threshold at which the micrographs are binarized by $\pm20\%$, as shown by the rows.}
    \label{figtableS2}
\end{figure}

\newpage\subsection*{Movie captions}
\noindent\textbf{Movie S1.} Extended deposition of amine-functionalized polystyrene particles injected at $260$ kPa. Black circles show cross-sections through the beads making up the porous media, white space shows pore space, and red shows deposited colloidal particles. For clarity, we only show the deposition profile for the first $4000~\upmu$m of the medium. Flow direction is from left to right.

\noindent\textbf{Movie S2.} Monotonic deposition of amine-functionalized polystyrene particles upstream of a bead during injection at $260$ kPa. Black circles show cross-sections through the beads making up the porous media, white space shows pore space, and red shows deposited colloidal particles. Flow direction is from left to right.

\noindent\textbf{Movie S3.} Cyclic deposition and erosion of amine-functionalized polystyrene particles upstream of a bead during injection at $260$ kPa. Black circles show cross-sections through the beads making up the porous media, white space shows pore space, and red shows deposited colloidal particles. Flow direction is from left to right.

\noindent\textbf{Movie S4.} Localized deposition of amine-functionalized polystyrene particles injected at $80$ kPa. Black circles show cross-sections through the beads making up the porous media, white space shows pore space, and red shows deposited colloidal particles. For clarity, we only show the deposition profile for the first $4000~\upmu$m of the medium. Flow direction is from left to right.

\noindent\textbf{Movie S5.} Extended deposition of carboxyl-functionalized polystyrene particles injected at $170$ kPa. Black circles show cross-sections through the beads making up the porous media, white space shows pore space, and red shows deposited colloidal particles. For clarity, we only show the deposition profile for the first $6000~\upmu$m of the medium. Flow direction is from left to right.

\noindent\textbf{Movie S6.} Monotonic deposition of carboxyl-functionalized polystyrene particles in a pore throat between beads during injection at $170$ kPa. Black circles show cross-sections through the beads making up the porous media, white space shows pore space, and red shows deposited colloidal particles. Flow direction is from left to right.

\noindent\textbf{Movie S7.} Cyclic deposition and erosion of carboxyl-functionalized polystyrene particles in a pore throat between beads during injection at $170$ kPa. Black circles show cross-sections through the beads making up the porous media, white space shows pore space, and red shows deposited colloidal particles. Flow direction is from left to right.

\noindent\textbf{Movie S8.} Localized deposition of carboxyl-functionalized polystyrene particles injected at $80$ kPa. Black circles show cross-sections through the beads making up the porous media, white space shows pore space, and red shows deposited colloidal particles. For clarity, we only show the deposition profile for the first $6000~\upmu$m of the medium. Flow direction is from left to right.

\end{document}